\documentstyle[12pt]{article}
\newcommand{\be}{\begin{equation}}
\newcommand{\ee}{\end{equation}}
\newcommand{\non}{\nonumber}
\newcommand{\bea}{\begin{eqnarray}}
\newcommand{\eea}{\end{eqnarray}}
\newcommand{\ba}{\begin{array}}
\newcommand{\ea}{\end{array}}
\newcommand{\al}{\alpha}
\newcommand{\pa}{\partial}

\newcommand{\si}{\sigma}
\newcommand{\la}{\lambda}
\newcommand{\ta}{\tau}
\newcommand{\ga}{\gamma}
\newcommand{\om}{\omega}
\newcommand{\de}{\delta}
\newcommand{\ze}{\zeta}
\newcommand{\ka}{\kappa}

\newcommand{\tha}{\theta}

\newcommand{\rar}{\rightarrow}

\newcounter{mycount}

\begin{document}

\begin{titlepage}
\vspace{-1mm}

\begin{flushright}
G\"{o}teborg ITP 97-05 \\ M\'exico ICN-UNAM 97-02
\end{flushright}
\vspace{1mm}
\begin{center}{\bf\Large\sf Hidden Algebras of the (super) Calogero
and Sutherland models}
\end{center}
\vskip 12mm
\begin{center}{{\bf\large Lars Brink,{\normalsize ${}^{\S}$
\footnote{tfelb@fy.chalmers.se}} Alexander Turbiner{\normalsize ${}^{\ddag}$
\footnote{turbiner@xochitl.nuclecu.unam.mx}${}^{,}
$\footnote{On leave of absence from the Institute for Theoretical
and Experimental Physics, \\ \indent \hspace{5pt} Moscow 117259, Russia.}}
and Niclas Wyllard{\normalsize ${}^{\S}$
\footnote{wyllard@fy.chalmers.se}}\vspace{8mm}}\\
${}^{\S}${\em  Institute of Theoretical Physics, S-412 96 G\"{o}teborg,
Sweden}} \\ \vspace{4mm}
${}^{\ddag}${\em Instituto de Ciencias Nucleares, UNAM,
Apartado Postal 70-543,\\ 04510 Mexico D.F., Mexico}
\end{center}

\vskip 18mm

\begin{abstract}
We propose to parametrize the configuration space of one-dimensional
quantum systems of $N$ identical particles by the elementary symmetric
polynomials of bosonic and fermionic coordinates.
It is shown that in this parametrization the Hamiltonians of the
$A_{N}$, $BC_N$, $B_{N}$, $C_{N}$ and $D_{N}$ Calogero and Sutherland models,
as well as their supersymmetric generalizations,
can be expressed --- for {\it arbitrary} values of the coupling constants ---
as quadratic polynomials in the generators of a Borel subalgebra of the Lie
algebra $gl(N+1)$
or the Lie superalgebra $gl(N+1|N)$ for the supersymmetric case.
These algebras are realized by first order differential operators.
This fact establishes the exact solvability of the models according to
the general definition given by one of the authors in 1994,
and implies that the Calogero and Jack-Sutherland polynomials, as well as
their supersymmetric generalizations, are related to finite-dimensional
irreducible representations of the Lie algebra $gl(N+1)$ and the Lie
superalgebra $gl(N+1|N)$.
\end{abstract}

\end{titlepage}

\setcounter{equation}{0}
\section{Introduction}

Exact solutions of non-trivial problems are always of great importance.
They give a hint about the structure of real problems and also provide a
laboratory
for testing approximate methods. The non-relativistic many-body
Calogero-Sutherland models \cite{Calogero,Sutherland:1971} together with their
supersymmetric extensions \cite{Freedman:1990,Brink:1993,Shastry:1993} provide
one of the most valuable sources  of exact solutions to one-dimensional
many-body quantum mechanical systems. The goal of this article is to try to
uncover a hidden reason
for the solvability of these models in order to answer the question what is
special about them and whether there are other exactly-solvable or
quasi-exactly-solvable many-body problems.

The remarkable discovery of the solvability of the bosonic $N$-body Calogero
\cite{Calogero} and Sutherland \cite{Sutherland:1971} models was at the time a
state-of-the-art achievement. Of course, this raised the question of whether
there existed a regular procedure to generate these models. After a few years,
it was found that the models are connected with the root systems of the
$A_{N-1}$ Lie algebras \cite{Olshanetskii:1977}, and can be obtained from
the Laplace-Beltrami operators defined on the symmetric spaces; this
procedure was called `the method of Hamiltonian reduction'
\cite{Olshanetskii:1977,Kazhdan:1978,Olshanetskii:1981,Olshanetskii:1983}
\footnote{For a contemporary viewpoint on the Hamiltonian reduction method see,
for example, \cite{Gorskii:1994}.}. The Hamiltonian reduction method provides a
regular basis for an explanation of the solvability of the Calogero and
Sutherland
models, at least for a selected set of values of the coupling constant(s).
Considerations of the root systems of the other simple Lie algebras made it
possible to find several additional families of exactly-solvable
multi-dimensional Schr\"odinger equations and to prove the complete
integrability of all these models.

Recently, another explanation of the exact-solvability of bosonic
many-body problems was presented, which was based on the finding that the
eigenfunctions of the $N$-body $A_{N-1}$ Calogero-Sutherland models form a
flag coinciding with the flag of the finite-dimensional representation spaces
of the Lie algebra $gl(N)$ \cite{Turbiner:1994,Ruhl:1995}. It was shown
that the Hamiltonians of the Calogero and Sutherland models are nothing but
different non-linear elements of the universal enveloping algebra of a Borel
subalgebra of the $gl(N)$ algebra. Unlike the ``method of reduction'',
in the second method the coupling constants appear as  certain parameters
fixing the element of the universal enveloping algebra. The second method
can be used to explain the solvability of the above models for {\it all}
allowed values of the coupling constant(s). In the present paper we show
that the $BC_{N}$, $B_N$, $C_{N}$ and $D_N$ Calogero (rational) and Sutherland
(trigonometric) Hamiltonians, for {\it any} values of the coupling constants,
are second order polynomials in the generators of the $gl(N+1)$ algebra.
These results were made possible by exploiting the fact that the configuration
space of the above quantum-mechanical systems can be parametrized by variables
in which the permutation symmetry is already encoded. The most suitable
variables are given by the elementary symmetric polynomials of the coordinates
of the particles. In these variables the eigenfunctions have an especially
simple form.

Other approaches have also been used to study the Calogero and Sutherland
models. Some of these have been directed towards obtaining closed
expressions for the wave functions. The bosonic $A_{N}$ Calogero model was
solved using an operator method in \cite{Brink:1992}, a similar method was
applied to the bosonic $A_{N}$ Sutherland model in \cite{Lapointe:1996}.
The eigenfunctions of the $BC_N$ Sutherland model has been studied in
\cite{Olshanetskii:1983,Bernard:1995} (ground-state) and \cite{Serban:1997}.
It is worth mentioning the remarkable observation that the wave functions
of the bosonic $A_N$ Sutherland models are in correspondence with singular
vectors of $W$-algebras (see, for example, \cite{Awata:1994}).

Quite recently, supersymmetric extensions of the many-body $A_{N}$ Calogero
\cite{Freedman:1990,Brink:1993} and Sutherland \cite{Shastry:1993} models
were constructed using the standard prescription of supersymmetrization.
In the present article we derive the supersymmetric extensions of the $BC_{N}$,
$B_N$, $C_{N}$ and $D_N$ Calogero and Sutherland models and
show that their solvability can be explained by the existence of the hidden
Lie superalgebra $gl(N+1|N)$.  It is worth mentioning that a supersymmetric
analogue of the Hamiltonian reduction method has not so far been constructed
for these models.

The paper is organized as follows: In the next Section we briefly review the
bosonic many-body $A_{N}$ Calogero and Sutherland models with emphasis on
the property of exact solvability, and also set up our notation. In Section 3
we study the supersymmetric extension of the Calogero model and show that it
is exactly solvable. In Section 4 the same analysis is carried out for the
supersymmetric Sutherland model. Section 5 is devoted to the Calogero models
connected with the Lie algebras $BC_{N}$, $B_N$, $C_{N}$ and $D_N$ as well as
their supersymmetric extensions; it is shown that those models are exactly
solvable.
In Section 6 the $BC_{N}$, $B_N$, $C_{N}$ and $D_{N}$ Sutherland models and
their
supersymmetric extensions are explored and their exact solvability
is established. The Conclusion contains a summary of the results obtained
and a discussion of some possible directions for future investigations.
In Appendix A we present a realization of the Lie algebra $gl(N)$ and the Lie
superalgebra
$gl(N|M)$ in terms of  first order differential operators.
Appendix B contains the Lie algebraic forms of the Hamiltonians
discussed in this paper. Finally, in Appendix C we give some details on the
derivation of the Lie algebraic forms of the Hamiltonians for the Calogero and
Sutherland models.

\setcounter{equation}{0}
\section{Bosonic (many-body) $A_{N-1}$ Calogero and Sutherland models
(review)}
\label{review}

In this section we briefly review, following \cite{Ruhl:1995}, the
{\it exact-solvability} of the bosonic many-body Calogero and Sutherland
models or, more precisely, the $A_{N-1}$ Calogero and Sutherland
models\footnote{For the sake of simplicity, in Sections 2-4 we refer to
these models as the Calogero and Sutherland models.}.

The Calogero and Sutherland models describe systems of $N$ identical
particles situated on the line and the circle, respectively.
The degrees of freedom of these models are parametrized by $N$ real coordinates
$x_i$.
The Hamiltonian of the Calogero model is defined by
\be
\label{e1.1}
        {\cal H}_{{\rm Cal}} = \frac{1}{2}\sum_{i=1}^{N}
\bigg[-\frac{\pa^{2}}{\pa x_{i}^{2}} + \om^2 x_{i}^{2}\bigg] +
g\sum_{i<j}\frac{1}{(x_{i}-x_{j})^{2}}\ ,
\ee
where $g=\nu(\nu -1) > -\frac{1}{4}$ is the coupling constant and
$\om$ is the harmonic oscillator frequency. For convenience only,
we place the center-of-mass under the influence of the harmonic oscillator
force. The ground state eigenfunction is given by
\be
\label{e1.2}
\Psi_{0}^{(c)}(x) = \Delta^{\nu}(x) e^{-\om\frac{X^{2}}{2}}\ ,
\ee
where $\Delta(x) = \prod_{i<j}|x_{i}-x_{j}|$ is the Vandermonde determinant
and $X^{2} = \sum_{i}x_{i}^{2}$.
\footnote{One should stress the point that to a fixed value of the coupling
constant $g$ there correspond two different values of the
parameter $\nu$, namely, $\nu=\al$ and $\nu=(1-\al)$, giving rise to two
families of eigenfunctions. Of course, $\al$ should be chosen in such a way
as to minimize the eigenvalue; then (\ref{e1.2}) corresponds to the ground
state. The value $\nu = (1-\al)$ inserted in (\ref{e1.2}) describes the ground
state (if normalizable) but of {\it another} family. If $g=0$, this
family comprises the states of negative parity with respect to permutations.}
As has been  shown by F.~Calogero,
any eigenstate of the system can be written in the form
\be
\label{e1.3}
\Psi(x)\ =\ \Psi_0^{(c)}(x) P_{c}(x)\,,
\ee
where $P_{c}(x)$ is a certain polynomial in the $x_i$'s, symmetric under the
permutation of any two particles. We refer to these polynomials as {\it
Calogero
polynomials}. The operator having the Calogero polynomials
as eigenfunctions is obtained from the Hamiltonian (\ref{e1.1}) by a
``gauge'' rotation, by which we mean the similarity transformation
\be
\label{e1.4}
h_{{\rm Cal}} = (\Psi_{0}^{(c)})^{-1}{\cal H}_{{\rm Cal}}\Psi_{0}^{(c)}\ .
\ee

The Hamiltonian of the Sutherland model is defined by
\be
\label{e1.5}
{\cal H}_{\rm Suth} = -\frac{1}{2}\sum_{k=1}^{N}\frac{\pa^{2}}{\pa x_{k}^{2}}
 + \frac{g}{4}\sum_{k<l}\frac{1}{\sin^{2}(\frac{1}{2}(x_{k} - x_{l}))}
\ee
where $g=\nu(\nu -1) > -\frac{1}{4}$ is the coupling constant\footnote{See
footnote 7.}.
The ground state eigenfunction is given by
\be
\label{e1.6}
\Psi_{0}^{(s)}(x) = (\Delta^{(trig)}(x))^{\nu}\ ,
\ee
where $\Delta^{(trig)}(x) = \prod_{i<j}|\sin(\frac{1}{2}(x_{i}-x_{j})|$
is a trigonometric analogue of the Vandermonde determinant.  It was shown
by B.~Sutherland that similarly to the Calogero model any eigenfunction of the
Hamiltonian (\ref{e1.5}) can be written in the form
\be
\label{e1.7}
\Psi(x)\ =\ \Psi_0^{(s)}(x) P_{s}(e^{ix})\,,
\ee
where $P_{s}(e^{ix})$ is a certain polynomial in $e^{ix_j}$, symmetric under
the permutation of any two particles. These polynomials are the so called
Jack-Sutherland polynomials \cite{Jack:1969,Sutherland:1971} (for a general
description, see, for example, \cite{Stanley:1989,MacDonald:1995}).
Performing a gauge rotation of ${\cal H}_{\rm Suth}$ with the gauge factor
$\Psi_0^{(s)}$ we arrive at the operator
\be
\label{e1.8}
h_{{\rm Suth}} = (\Psi_{0}^{(s)})^{-1}{\cal H}_{{\rm Suth}}\Psi_{0}^{(s)}\ ,
\ee
which has the Jack-Sutherland polynomials as eigenfunctions.

In order to study the internal dynamics we
introduce the center-of-mass coordinate $Y\ =\ \sum_{j=1}^N x_j$ and
the translation-invariant relative Perelomov coordinates \cite{Perelomov:1971}:
\be
\label{e1.9}
y_i\ =\ x_i - \frac{1}{N} Y\ ,\quad i=1,2,\ldots , N \quad ,
\ee
which obey the constraint $\sum_{i=1}^N y_i=0$. In order to incorporate the
permutation symmetry and the translation invariance we consider two sets
of coordinates \cite{Ruhl:1995}:
\be
\label{e1.10}
(x_1,x_2,\ldots x_N) \rightarrow (Y, \tau_n (x)=\si_n(y(x))|
\ n=2,3,\ldots , N)\,,
\ee
and
\be
\label{e1.11}
(x_1,x_2,\ldots x_N) \rightarrow (e^{iY}, \eta_n (x)=\si_n(e^{iy(x)})|
\ n=1,2,\ldots , (N-1))\,,
\ee
where
\be
\si_{k}(x) = \sum_{i_{1}<i_{2}<\cdots<i_{k}} x_{i_{1}}x_{i_{2}}\cdots x_{i_{k}}
\ee
are the elementary symmetric polynomials (see, for example,
\cite{MacDonald:1995}). $\si_{k}(y(x))$ are thus the elementary symmetric
polynomials with translation-invariant arguments\footnote{Due to the constraint
$\sum_{i=1}^N y_i=0$, the symmetric
polynomials can be considered as defined in ${\bf R}^N$ and then restricted
to the hyperplane $\sum_{i=1}^N y_i=0$.}.
Then the following statement holds \cite{Ruhl:1995}:
\begin{quote}
{\it After separation of the center-of-mass,
the operators $h_{{\rm Cal}}$ and $h_{{\rm Suth}}$, when written in the
coordinates $\tau$ (\ref{e1.10}) and $\eta$ (\ref{e1.11}), respectively,
are quadratic polynomials in the generators of the Borel subalgebra of
$gl(N)$ spanned by the operators $J^{0}_{ij}$ and $J^{-}_{i}$ realized as first
order differential operators (see Appendix A, eq. (\ref{a1})):
\[
h = \sum_{i,j,k,l=2}^{N} A_{ijkl}J^{0}_{ij}J^{0}_{kl} +
\sum_{i,j,k=2}^{N}B_{ijk} J^{0}_{ij}J^{-}_{k} +
\sum_{i,j=2}^{N}C_{ij}J^{0}_{ij} + \sum_{i=2}^{N}D_{i}J^{-}_{i} \,,
\]
(see Appendix B). The operators $J^{0}_{ij}$ and $J^{-}_{i}$  can be
represented by triangular matrices and preserve the flag of spaces of
inhomogeneous polynomials
\[
\label{e1.12}
{\cal P}_n \ =\ \mbox{\rm span} \{ v_2^{n_2} v_3^{n_3} v_4^{n_4}
\ldots v_{N}^{n_{N}} |0 \leq \sum n_i \leq n \}\ ,
\]
where $v_k=\tau_k$ and  $v_k=\eta_{k+1}$, respectively, and $k=2,\ldots N$.
The coupling constants $g$ (see (\ref{e1.1}) and (\ref{e1.5})) appear only
in the coefficients $C_{ij}$ and $D_{i}$ and are not related to the dimension
of the representation of $gl(N)$. Consequently, each Hamiltonian
(\ref{e1.1}), (\ref{e1.5}) has one or several infinite families of polynomial
eigenfunctions.}
\end{quote}
This statement leads to the important conclusion that the Calogero and
Jack-Sutherland polynomials are intimately connected with the
finite-dimensional irreducible representations of the Lie algebra $gl(N)$.
It also provides a simple computational tool for deriving the explicit form
of the Calogero and Jack-Sutherland polynomials.

\setcounter{equation}{0}
\section{The supersymmetric many-body Calogero model}
\label{sCal}

In order to proceed to the problem of the supersymmetric generalizations
of (\ref{e1.1}) and (\ref{e1.5}), let us first recall \cite{Freedman:1990}
that in a supersymmetric system of particles each bosonic degree of
freedom $x_i$ is accompanied by the fermionic variables $\vartheta_i$ and
$\vartheta_i^{\dagger}$, which obey the standard anti-commutation rules
$\{\vartheta_i,\vartheta_j\}\ =
\{\vartheta_i^{\dagger},\vartheta_j^{\dagger}\}\ =0$ and
$\{\vartheta_i,\vartheta_j^{\dagger}\} =\ \delta_{ij}$.  Throughout the paper
we use a concrete realization of this algebra:
\[
\vartheta_i=\tha_i\,,\ \; \vartheta_i^{\dagger}\ =\ \frac{\pa}{\pa \tha^{i}} \
{}.
\]
It is convenient to introduce the `cmino' $\Psi$ --- the fermionic analogue of
the center-of-mass coordinate as
\be
\label{e1.13}
\Psi\ =\ \sum_{i=1}^N \tha_i\,,
\ee
which is the super-partner of $Y$. We can also define the fermionic analogue
of the Perelomov coordinates (\ref{e1.9})
\be
\label{e1.14}
\la_i=\tha_i\ -\ \frac{1}{N}\Psi\,.
\ee
In order to construct the supersymmetric many-body Calogero model let us
introduce the supercharges $Q$ and $Q^{\dagger}$ as defined in
\cite{Freedman:1990}:
\be
\label{q2.1}
Q=\sum_k \frac{\pa}{\pa\tha_{k}}\bigg(p_k-i\frac{\pa W}{\pa x^{k}}\bigg) \ ,\
Q^{\dagger}=\sum_k \tha_{k} \bigg(p_k+i\frac{\pa W}{\pa x^{k}}\bigg)\ ,
\ee
where $Q^2={Q^{\dagger}}^2=0$, $p_k=-i\frac{\pa}{\pa x_k}$, and the
superpotential is
\be
\label{q2.2}
W(x_1,x_2,\ldots,x_N)= -\frac{\om}{2} \sum_{i=1}^{N}x_i^2 +
\nu\sum_{i<j}\log|x_i-x_j|  \ .
\ee
Then the supersymmetric Hamiltonian ${\cal H}_{\rm sCal} =
\frac{1}{2}\{Q,Q^{\dagger}\}$ has the form  \cite{Freedman:1990}
\be
\label{e2.1}
{\cal H}_{\rm sCal} =  \frac{1}{2}\sum_{i=1}^{N}\bigg[-\frac{\pa^{2}}{\pa
x_{i}^{2}} + \om^2 x_{i}^{2}\bigg] +
\om\sum_{i=1}^N\tha_{i}\frac{\pa}{\pa \tha^{i}} +
\sum_{i<j}\frac{\nu}{(x_{i}-x_{j})^{2}}\bigg[(\nu-1) +
\tha_{i}-\tha_{j}(\frac{\pa}{\pa\tha_{i}} - \frac{\pa}{\pa\tha_{j}})\bigg] +
C \ ,
\ee
where $C = -\frac{1}{2}\nu\om N(N-1) - \frac{1}{2}N\om$.
The ground state eigenfunction remains the same (cf. (\ref{e1.2}))
as for the bosonic many-body Calogero model (\ref{e1.1}).
It is easy to see that a gauge rotation of ${\cal H}_{\rm sCal}$ with the
ground
state wave function $\Psi_0^{(c)}$ as a gauge factor affects only the bosonic
part of the Hamiltonian. Defining
\be
h_{\rm sCal}=-2(\Psi_0^{(c)})^{-1}{\cal H}_{\rm sCal}\Psi_0^{(c)}
\ee
(cf. (\ref{e1.4})), we obtain
\be
\label{e2.2}
h_{\rm sCal} = \sum_{i=1}^{N}\frac{\pa^{2}}{\pa x_{i}^{2}}
-2\om\sum_{i=1}^{N}[x_{i}\frac{\pa}{\pa x_{i}} +\tha_{i}\frac{\pa}{\pa
\tha^{i}}]
+ 2\nu\sum_{i<
j}[\frac{1}{x_{i}-x_{j}}(\frac{\pa}{\pa x_{i}} - \frac{\pa}{\pa x_{j}}) -
\frac{\tha_{i}-\tha_{j}}{(x_{i}-x_{j})^{2}}(\frac{\pa}{\pa\tha_{i}} -
\frac{\pa}{\pa\tha_{j}})].
\ee
This expression can be called the {\it rational} form of the supersymmetric
many-body Calogero model.
Similarly to the bosonic case the eigenfunctions of the operator
(\ref{e2.2}) after separation of the center-of-mass remain
polynomials but now in $(y^{i},\lambda^{i})$. These polynomials are symmetric
under
the permutation of any pair of particles
$(x^{i},\tha^{i}) \leftrightarrow (x^{j},\tha^{j})$, and
can be considered as the supersymmetric analogue of the Calogero polynomials.

Now let us introduce permutation-symmetric variables. One can construct
two sets of such variables: (i) a set analogous to the Newton
polynomials
\be
\label{e2.3}
s_{n} = \sum_{i=1}^{N} x_{i}^{n} \quad ,\quad \rho_{n} = \sum_{i=1}^{N}
\tha_{i}x_{i}^{n-1}
\ee
and (ii) a set analogous to the elementary symmetric polynomials
\be
\label{e2.4}
\si_{k} = \sum_{i_{1}<i_{2}<\cdots<i_{k}} x_{i_{1}}x_{i_{2}}\cdots x_{i_{k}}
\quad ,\quad \ze_{k} = \frac{1}{(k-1)!}\sum_{i_{1}\neq i_{2}\neq\cdots\neq
i_{k}}\tha_{i_{1}}x_{i_{2}}\cdots x_{i_{k}} \,.
\ee
The variables $s_{n}$ and $\rho_{n}$ as well as $\si_{k}$ and $\ze_{k}$
are symmetric under the permutation,
$(x^{i},\tha^{i}) \leftrightarrow (x^{j},\tha^{j})$. However, the property
$\si_{N+k} = 0\ ,\ \ze_{N+k} = 0\ ,\ k=1,2,\ldots$ implies that the
variables $\si_{k}$ and $\ze_{k}$, has the advantage that they
avoid the difficulties associated with the overcompleteness of the basis,
which plague the variables $s_{n}$ and $\rho_{n}$. Therefore,
in the sequel we make use of the variables $\si_{k}$ and $\ze_{k}$.

It is worth mentioning the following relations between the two sets
of variables
\bea
\label{e2.5}
\sum_{k=0}^{N} \si_{k}(x)t^{k} &=& \exp
[\sum_{n=1}^{\infty}\frac{(-1)^{n+1}}{n}s_{n}(x)t^{n} ]\,, \non \\
\sum_{k=0}^{N}
\ze_{k}(x)t^{k} &=& (\sum_{m=1}^{\infty} (-1)^{m+1}\rho_{m}(x,\tha)t^{m})\exp
[\sum_{n=1}^{\infty}\frac{(-1)^{n+1}}{n}s_{n}(x)t^{n} ]\,.
\eea
A convenient way of succinctly writing these relations is obtained if one
introduces the ``superspace'' coordinates
\bea
\label{ss-coord}
\chi_{i} = \si_{i} + \al\ze_{i}\,, \;\; \phi_{i} = x_{i} + \al\tha_{i}\,,
\eea
 where $\al$ is a Grassmann number, $\al^{2} = 0$; then
\bea
\label{e2.6}
\sum_{k=0}^{N} \chi_{k}(\phi_{i})t^{k} &=& \exp
[\sum_{n=1}^{\infty}\frac{(-1)^{n+1}}{n}s_{n}(\phi_{i})t^{n} ]\,,
\eea
encodes both the relations (\ref{e2.5}).

The ``superspace'' formulation makes it possible to write formulas in a more
compact way. The ``supercoordinate'' $\phi_{i}$ is clearly an $N=1$ superfield.
However, one should stress that the models we study possess
$N=2$ supersymmetry, while the ``superspace'' formulation is not manifestly
$N=2$ supersymmetric. Nevertheless, later on we will see explicitly
that the superspace (we drop the quotation marks hereafter) formulation
turns out to be a useful computational aid.
For example, one can rewrite the Hamiltonian $h_{\rm sCal}$ in
the superspace coordinates as follows
\be
\label{e2.7}
h_{\rm sCal} = \int d\al d\bar{\al} \sum_{i=1}^{N}
\frac{\de}{\de\bar{\phi}_i}\frac{\de}{\de\phi_{i}} -
\int d\al {\cal W} - \int d\bar{\al}\bar{{\cal W}} \ .\label{susyham}
\ee
Here $\bar{\al}$ is a Grassmann variable independent of $\al$ and
\be
\label{ss-der}
\frac{\de}{\de\phi_{i}} =
\frac{\pa}{\pa\tha_{i}} + \al\frac{\pa}{\pa x_{i}}\ , \quad
\frac{\de}{\de\bar{\phi}_i} =
\frac{\pa}{\pa\tha_{i}} + \bar{\al}\frac{\pa}{\pa x_{i}}\ ,
\ee
while
\be
\label{e2.8}
 {\cal W} = \om\sum_{i=1}^{N}\phi_{i}\frac{\de}{\de\phi_{i}} -
\nu\sum_{i<j}\frac{1}{\phi_{i}-\phi_{j}}
(\frac{\de}{\de\phi_{i}} - \frac{\de}{\de\phi_j}) \ ,\quad
\bar{{\cal W}} = {\cal W}(\bar{\phi})\ ,
\ee
where $\bar{\phi}_{i} = x_{i} + \bar{\al}\tha_{i}$. Notice that we have
not introduced any superderivatives; $\frac{\de}{\de\phi_{i}}$ and
$\frac{\de}{\de\bar{\phi}_i}$ should be considered as ``superfields''
(since they do not contain derivatives with respect to $t$, $\al$ and/or
$\bar{\al}$).

Like we did in Section 2 we would like to separate the dynamics of the
center-of-mass and the relative motion, and encode in relative coordinates
the translation invariance and permutation symmetry of the system. In order
to implement these requirements we introduce the translation-invariant and
permutation-symmetric coordinates of the relative motion
\be
\label{ss-cal}
\ta_{k}(y) = \si_{k}(y(x))\, ,\quad \ka_{k}(y,\la) = \ze_{k}(y(x),\la(\tha))\,,
\ee
for example, in the symmetric polynomials (\ref{e2.4}) the arguments ($x$'s)
are replaced by the
translation-invariant $y$'s (\ref{e1.9}) and the $\tha$'s are replaced
by the translation-invariant $\la$'s (\ref{e1.14}). Here $\ta_0 = 1,\ \ta_1=0$
and $\ka_0 = \ka_1 = 0$. The next step is to make the change of variables
\be
\label{e2.9}
(x_{i},\tha_{i} |i=1,2,\ldots,N) \rightarrow
(Y,\ta_{k};\Psi,\ka_{k} |\ k=2,3,\ldots,N)\ .
\ee
in the Hamiltonian $h_{\rm sCal}$. In general, this is a tedious and cumbersome
calculation. Fortunately, the superspace formulation allows us to avoid
most of the tiresome algebraic calculations. Let us define
\be
\tilde{\phi}_{i} = \phi_{i} - \frac{1}{N}\sum_{j=1}^{N}\phi_{j}\,.
\ee
Then the expression for $\psi_{i} = \tau_{i} + \al\ka_{i}$ can be obtained
from the relation
\be
\sum_{k=0}^{N}\psi_{k}(\tilde{\phi}_{i})t^{k} =
\exp[\sum_{n=1}^{\infty}\frac{(-1)^{n}}{n}s_{n}(\tilde{\phi}_{i})t^{n}]\,.
\ee
The derivative $\frac{\de}{\de \phi_{i}(\al)} =
 \frac{\pa}{\pa \tha_{i}} + \al\frac{\pa}{\pa x_{i}}$  satisfies
\be
\frac{\de \phi_{j}(\beta)}{\de \phi_{i}(\al)} = \de_{ij}(\al - \beta)
 = \de_{ij}\de (\al - \beta)\,,
\ee
where we have used that $\int d\al \de(\al - \beta)f(\al) = f(\beta)$. We now
make the change of variables
\[
(\phi_{i}|i=1,\ldots N) \rar (\chi_{1} =
 \sum_{k=1}^{N}\phi_{k}\,;\ \psi_{j}|j=2,\ldots N)\ .
\]
Under such a change of variables the derivatives transform
as\footnote{From now on we suppress the center-of-mass coordinate $\chi_{1}$.}
\be
\label{e3.20}
\frac{\de}{\de \phi_{i}(\al)} = - \int d\beta
 \frac{\de \psi_{j}(\beta)}{\de \phi_{i}(\al)}
\frac{\de}{\de \psi_{j}(\beta)}\,,
\ee
where $\frac{\de}{\de \psi_{i}(\al)} = \frac{\pa}{\pa \ka_{i}}
+ \al\frac{\pa}{\pa \tau_{i}}$, and summation over $j$ is assumed.
To prove the relation (\ref{e3.20}) one simply writes both sides of the
equality in components. It is convenient to perform the change
of variables in two steps: to write,
\be
\frac{\de}{\de \phi_{i}(\al)} =\int d\beta d\ga \frac{\de
\tilde{\phi}_{j}(\ga)}{\de \phi_{i}(\al)}\frac{\de
\psi_{k}(\beta)}{\de \tilde{\phi}_{j}(\ga)}
\frac{\de}{\de \psi_{k}(\beta)}\,,
\ee
and then use the definition of $\tilde{\phi_{i}}$ together with (\ref{e3.20})
which leads to $\frac{\de \tilde{\phi}_{j}(\ga)}{\de \phi_{i}(\al)} =
(\de_{ij} - \frac{1}{N})\de(\al - \ga)$.
Let us give as an example the expression for the Laplace operator in the new
coordinates
\bea
&&\sum_{i=1}^{N} \int d\al d\beta \frac{\de}{\de \phi_{i}(\beta)}
\frac{\de}{\de \phi_{i}(\al)} \rar \sum_{i=1}^{N}
\sum_{j,k=2}^{N}\sum_{l,m=1}^{N} \int d\al d\beta \nonumber
 \\ && \int d\ga (\de_{il} - \frac{1}{N})
\frac{\de \psi_{j}(\ga)}{\de \tilde{\phi}_{l}(\beta)}
\frac{\de}{\de \psi_{j}(\ga)}\int d\epsilon (\de_{im} -
 \frac{1}{N})\frac{\de \psi_{k}(\epsilon)}{\de
 \tilde{\phi}_{m}(\al)}\frac{\de}{\de \psi_{k}(\epsilon)}\,.
\eea
The above considerations imply that within the framework of the superspace
formalism the calculations for the supersymmetric models follow closely
the calculations carried out in the bosonic case \cite{Ruhl:1995}.
In fact, for most of the terms the supersymmetric results can be obtained from
the bosonic ones by judiciously replacing the bosonic coordinates by
superspace coordinates. More details can be found in Appendix \ref{appC}. The
derivation makes extensive use of the generating function (\ref{e2.6}).

In the superspace coordinates the Laplace operator becomes (after reinserting
the center-of-mass coordinate)
\bea
\label{}
\sum_{k=1}^{N}\int d\al d\bar{\al}
\frac{\de}{\de\bar{\phi}_{k}}\frac{\de}{\de\phi_{k}} &=& \int d\al d\bar{\al}
\left[ N\frac{\de}{\de \bar{\chi}_{1}}\frac{\de}{\de \chi_{1}} +
\sum_{i,j=2}^{N}{\cal A}_{ij}
\frac{\de}{\de \bar{\psi}_{j}}\frac{\de}{\de \psi_{i}}\right] \non \\ && + \int
d\al \sum_{i=2}^{N}{\cal B}_{i}\frac{\de}{\de \psi_{i}}  + \int d\bar{\al}
\sum_{i=2}^{N}\bar{{\cal B}}_{i}\frac{\de}{\de \bar{\psi}_{i}}\,,
\eea
where
\be
\label{psi-coord}
\psi_i = \ta_{i} + \al\ka_{i} \,, \ \;\;
\frac{\de}{\de \psi_{i}} =
\frac{\pa}{\pa\ka_{i}} + \al\frac{\pa}{\pa \ta_{i}} \,,
\ee
and
\bea
\label{Aij}
{\cal A}_{ij} &=& \frac{(N-i+1)(j-1)}{N}\psi_{i-1}\bar{\psi}_{j-1} +
\sum_{l\geq\max (1,j-i)}(j-i-2l)\psi_{i+l-1}\bar{\psi}_{j-l-1} \\ &-&
\sum_{l\geq 1}l\big[\bar{\psi}_{i+l}\psi_{j-2-l} -
\psi_{i+l}\bar{\psi}_{j-2-l}\big]\, , \non \\
{\cal B}_{i}(\psi) &=&
- \frac{1}{2N}\sum_{i=2}^{N}(N-i+2)(N-i+1)\psi_{i-2}\;\;\;,
\ \bar{\cal B}_{i} = {\cal B}_{i}(\bar{\psi})\, .
\eea
The final expression for the Laplace operator in components is given by
\bea
\label{e2.10}
\sum_{i=1}^{N} \frac{\pa^{2}}{\pa x_{i}^{2}} &=& N\frac{\pa^{2}}{\pa
Y^{2}} + \sum_{i,j=2}^{N}\bigg[A^{\ta\ta}_{ij}\frac{\pa^{2}}{\pa \ta_i\pa
\ta_j}
+ A^{\ta\ka}_{ij}\frac{\pa^{2}}{\pa \ta_i\pa\ka_j} +  A^{\ka \ta}_{ij}
\frac{\pa^{2}}{\pa \ta_j\pa\ka_i} + A^{\ka\ka}_{ij}\frac{\pa^{2}}{\pa
\ka_{j}\pa\ka_{i}}\bigg]  \non \\ &+& \sum_{i=2}^{N}
\bigg[B^{\ta}_{i}\frac{\pa}{\pa \ta_{i}}
+ B^{\ka}_{i}\frac{\pa}{\pa \ka_{i}}\bigg] \ ,
\eea
where
\bea
\label{e2.11}
A^{\ta\ta}_{ij} &=& \frac{(N-i+1)(j-1)}{N}\ta_{i-1}\ta_{j-1} + \sum_{l\geq\max
(1,j-i)}(j-i-2l)\ta_{i+l-1}\ta_{j-l-1}\ , \non \\
A^{\ta\ka}_{ij} &=&  \frac{(N-i+1)(j-1)}{N}\ta_{i-1}\ka_{j-1} + \sum_{l\geq\max
(1,j-i)}(j-i-2l)\ta_{i+l-1}\ka_{j-l-1} \non \\&-& \sum_{l\geq
1}l\big[\ka_{i+l}\tau_{j-2-l} - \tau_{i+l}\ka_{j-2-l}\big]\ , \non \\
A^{\ka \ta}_{ij} &=& \frac{(N-i+1)(j-1)}{N}\ka_{i-1}\ta_{j-1} + \sum_{l\geq\max
(1,j-i)}(j-i-2l)\ka_{i+l-1}\ta_{j-l-1} \non \\&+& \sum_{l\geq
1}l\big[\ka_{i+l}\tau_{j-2-l} - \tau_{i+l}\ka_{j-2-l}\big] \ ,\non \\
A^{\ka\ka}_{ij} &=& \frac{(N-i+1)(j-1)}{N}\ka_{i-1}\ka_{j-1} + \sum_{l\geq\max
(1,j-i)}(j-i-2l)\ka_{i+l-1}\ka_{j-l-1} \cr
&+& 2\sum_{l\geq 1}l\ka_{i+l}\ka_{j-2-l}\ ,\non \\
B^{\ta}_{i} &=&  - \frac{(N-i+2)(N-i+1)}{N}\ta_{i-2}\ \non \\
B^{\ka}_{i} &=&  - \frac{(N-i+2)(N-i+1)}{N}\ka_{i-2}\,.
\eea
As expected, $A^{\ta\ta}_{ij}$ and $B^{\ta}_{i}$ coincide with the expressions
found in \cite{Ruhl:1995} for the bosonic Calogero model. Let us note
that the coefficient functions $A_{ij}$ and $B_{i}$ are second and first
order polynomials in $\tau_{i}$ and $\ka_{i}$, respectively.

Dropping the decoupled center-of-mass terms one can show that in superspace
coordinates the remaining part of the Hamiltonian $h^{\rm (rel)}_{\rm sCal}$,
which describes the relative motion, becomes
\be
\label{e2.13}
h_{\rm sCal}^{\rm (rel)} = \int d\al d\bar{\al}\sum_{i,j=2}^{N}{\cal
A}_{ij}\frac{\de}{\de \bar{\psi}_{j}}\frac{\de}{\de \psi_{i}} -
\int d\al {\cal W} - \int d\bar{\al}\bar{{\cal W}} \ ,
\ee
where ${\cal A}_{ij}$ is given in (\ref{Aij}) and
\be
\label{e2.14}
{\cal W} = \om\sum_{i=2}^{N}i\psi_{i}\frac{\de}{\de \psi_{i}} +
\frac{1}{2}(\frac{1}{N} +
\nu)\sum_{i=2}^{N}(N-i+2)(N-i+1)\psi_{i-2}\frac{\de}{\de \psi_{i}}\ ,\
\bar{\cal W} = {\cal W}(\bar{\psi})\,.
\ee
Finally, in component form $h_{\rm sCal}^{\rm (rel)}$ looks like
\bea
\label{e2.12}
h_{\rm sCal}^{\rm (rel)} &=&  \sum_{i,j=2}^{N}\bigg[A^{\ta\ta}_{ij}
\frac{\pa^{2}}{\pa \ta_i\pa\ta_j} + A^{\ta\ka}_{ij}
\frac{\pa^{2}}{\pa \ta_i\pa\ka_j} + A^{\ka\ta}_{ij}
\frac{\pa^{2}}{\pa \ka_i\pa\ta_j} + A^{\ka\ka}_{ij}
\frac{\pa^{2}}{\pa \ka_j\pa\ka_i}\bigg] \non \\ && -
(\frac{1}{N} + \nu)
\sum_{i=2}^{N}(N-i+2)(N-i+1)\bigg[\ta_{i-2}\frac{\pa}{\pa \ta_i} +
\ka_{i-2}\frac{\pa}{\pa \ka_i}\bigg] \non \\ && -
2\om \sum_{i=2}^{N}
i\bigg[\ta_i\frac{\pa}{\pa \ta_i} + \ka_i\frac{\pa}{\pa \ka_i}\bigg] \,.
\label{hrel}
\eea
where  $A^{\ta\ta}_{ij}, A^{\ta\ka}_{ij}, A^{\ka \ta}_{ij}, A^{\ka\ka}_{ij},
B^{\ta}_{i}, B^{\ka}_{i}$
are given in (\ref{e2.11}). This form can be called the {\it algebraic} form
of the supersymmetric many-body Calogero Hamiltonian.

It turns out that we are able to rewrite the supersymmetric Calogero
Hamiltonian $h_{\rm sCal}^{\rm (rel)}$ in terms of generators of the
Borel subalgebra of the algebra $gl(N|N-1)$ spanned by first order
differential operators in the variables $(\ta_{k},\ka_{k}),\ k=2,\ldots N$
(see Appendix A.2). The result is given in Appendix B, eq. (\ref{e2.16}).
Since the expression (\ref{e2.16}) contains no positive-root generators
$J^{+}_i, Q^+_{\ga}$, then in accordance with the general definition given in
\cite{Turbiner:1994} we conclude that the supersymmetric many-body
Calogero model (\ref{e2.1}) is exactly solvable. The existence of the
representation (\ref{e2.16}) proves that there are infinitely-many
eigenfunctions of the operator (3.7) having the form of polynomials in
the variables $(\ta_{k},\ka_{k})$. It also implies that eigenfunctions
of the supersymmetric many-body Calogero model (\ref{e2.1}) have
a factorizable form being the product of the ground-state eigenfunction
multiplied by a polynomial in the variables $(\ta_{k},\ka_{k})$. These
polynomials are related to finite-dimensional irreducible representations
of the algebra $gl(N|N-1)$ in the realization (A.3).

\setcounter{equation}{0}
\section{\hspace{-2mm}The supersymmetric many-body Sutherland model}
\label{sSuth}

In this section we analyze the exact-solvability of the
supersymmetric Sutherland model.

In order to define the supersymmetric extension of the bosonic many-body
Sutherland model (\ref{e1.5}) we use the standard prescription of
supersymmetrization already used in the previous Section.
Let us take the supercharges $Q, Q^{\dagger}$ (see (\ref{q2.1}))
with a superpotential
\be
\label{q3.2}
W(x_1,x_2,\ldots,x_N)= \nu\sum_{i<j}\log|\sin(\frac{1}{2}(x_i-x_j))|= \nu
\log\Delta^{(trig)}\,,
\ee
(cf. (\ref{q2.2})) and then construct the supersymmetric Hamiltonian
${\cal H}_{\rm sCal} = \frac{1}{2}\{Q,Q^{\dagger}\}$.
After carrying out the calculation the Hamiltonian of the supersymmetric
many-body Sutherland model emerges \cite{Shastry:1993}:
\be
\label{e3.1}
{\cal H}_{\rm sSuth} = -\frac{1}{2}\sum_{k=1}^{N}\frac{\pa^{2}}{\pa x_{k}^{2}}
 + \frac{1}{4}\sum_{k<l}\frac{\nu}{\sin^{2}(\frac{1}{2}(x_{k} - x_{l}))}
\bigg[\nu -
1 + (\tha_{k} - \tha_{l})(\frac{\pa}{\pa\tha_{k}} -
\frac{\pa}{\pa\tha_{l}})\bigg] + C\ ,
\ee
where $C = -N(N^{2} - 1)\nu^{2}/3$.
The ground state wave function remains the same as in the bosonic case,
$\Psi_{0}^{(s)} = {(\Delta^{(trig)}(x))}^{\nu}$, where
$\Delta^{(trig)}(x) = \prod_{k<l}|\sin(\frac{1}{2}(x_{k}-x_{l})|$.
Introducing the gauge-rotated Hamiltonian
$h_{\rm sSuth} = -2(\Psi^{(s)}_{0})^{-1}{\cal H}\Psi_{0}^{(s)}$, we get
\be
\label{e3.2}
h_{\rm sSuth} = \sum_{k=1}^{N}\frac{\pa^{2}}{\pa x_{k}^{2}} + \nu
\sum_{k<l}\bigg[ \cot(\frac{1}{2}(x_{k}-x_{l}))(\frac{\pa}{\pa x_{k}} -
\frac{\pa}{\pa x_{l}}) - \frac{1}{2}
\frac{(\tha_{k} - \tha_{l})}{\sin^{2}(\frac{1}{2}(x_{k} - x_{l}))}
(\frac{\pa}{\pa\tha_{k}} - \frac{\pa}{\pa\tha_{l}})\bigg]\,.
\ee
It is worth mentioning that if the operator (\ref{e3.2}) is written in the
coordinates $e^{ix_j}$, it appears in its {\it rational} form,
which is a supersymmetric generalization of the rational form of the Sutherland
Hamiltonian (see \cite{Sutherland:1971}).

One can show that like in the bosonic case (cf.(\ref{e1.8})) after separation
of the center-of-mass motion the eigenfunctions of the operator (\ref{e3.2})
remain polynomials but now in the coordinates $(e^{iy_{k}},\lambda_{k})$.
These polynomials are symmetric under the permutation of any pair of particles
$(x_{i},\tha_{i}) \leftrightarrow (x_{j},\tha_{j})$ and
can be considered as a supersymmetric analogue of the Jack-Sutherland
polynomials.

In the superspace coordinates (\ref{ss-coord}), (\ref{ss-der})
the Hamiltonian $h_{\rm sSuth}$ can be written
\be
\label{e3.3}
h_{\rm sSuth} = \int d\al
d\bar{\al}\sum_{k=1}^{N}\frac{\de}{\de\bar{\phi}_{i}}\frac{\de}{\de \phi_{i}} +
 \nu\int d\al {\cal W}  +  \nu\int d\bar{\al}\bar{{\cal W}}
\,,
\ee
where
\be
{\cal W}(\phi) = \frac{1}{2}\sum_{k<l}\cot(\frac{1}{2}(\phi_{k} -
\phi_{l}))(\frac{\de}{\de \phi_{k}} - \frac{\de}{\de \phi_{l}})\ ,\
\bar{{\cal W}} ={\cal W}(\bar{\phi})\,.
\ee
Next, we introduce the new variables (cf.(\ref{ss-cal}))
\be
\label{e3.4}
\eta_{n} + \alpha\mu_{n} = \si_{n}(\exp[i(y_{k} + \alpha\la_{k}) ])\ ,\
n = 1, \ldots, N-1  \quad ,
\ee
and
\be
\label{e3.5}
\si_{N} + \al\zeta_{N} = \si_{N}(\exp[i(x_{k} + \alpha\tha_{k}) ])\
,\ n = N\ ,
\ee
where $\al^{2} = 0$, $\si_{n}$ is defined in (\ref{e2.4}) and
$y_{i}, \la_{i}$ are given in (\ref{e1.9}), (\ref{e1.14}), respectively.
Furthermore,
\[
\sum_{k=0}^{N}\eta_{k}t^{k} =
\exp[\sum_{n=1}^{\infty}\frac{(-1)^{n+1}}{n}s_{n}(e^{i y_{k}})t^{n}]\ ,
\]
\be
\label{e3.6}
\sum_{k=0}^{N}\mu_{k}t^{k} =
(\sum_{m=1}^{\infty}(-1)^{m+1}\rho_{m}(e^{iy_{k}},
\la_{i}e^{iy_{k}})t^{m})\exp[\sum_{n=1}^{\infty}\frac{(-1)^{n+1}}{n}s_{n}
(e^{i y_{k}})t^{n}]\ ,
\ee
(cf.(\ref{e2.5})), where we have set $\eta_N=1$ and $\mu_N=0$.

Following the same procedure as in Section 3 one can rewrite the Laplace
operator in the new coordinates (\ref{e3.4}), (\ref{e3.5}):
\bea
\label{e3.8}
-\sum_{i=1}^{N} \frac{\pa^{2}}{\pa x_{i}^{2}} &\hspace{-3pt}=&\hspace{-3pt}
N(\si_{N}
\frac{\pa}{\pa\si_{N}})^{2} + \sum_{i,j=1}^{N-1}\bigg[A^{\eta\eta}_{ij}
\frac{\pa^{2}}{\pa \eta_{i}\pa \eta_{j}} +
A^{\eta\mu}_{ij}\frac{\pa^{2}}{\pa \eta_{i}\pa\mu_{j}} +
A^{\mu\eta}_{ij}\frac{\pa^{2}}{\pa \eta_{j}\pa\mu_{i}} +
A^{\mu\mu}_{ij}\frac{\pa^{2}}{\pa\mu_{j}\pa\mu_{i}}\bigg] \non \\ &+&
\sum_{i=1}^{N-1}\bigg[B^{\eta}_{i}\frac{\pa}{\pa\eta_{i}} +
B^{\mu}_{i}\frac{\pa}{\pa \mu_{i}}\bigg]  \ ,
\eea
where
\bea
\label{e3.9}
A^{\eta\eta}_{ij} &=& \frac{j(N-i)}{N}\eta_{i}\eta_{j} + \sum_{l\geq\max
(1,j-i)}(j-i-2l)\eta_{i+l}\eta_{j-l}\ , \non \\
A^{\eta\mu}_{ij} &=& \frac{j(N-i)}{N}\eta_{i}\mu_{j} +
\hspace{-4mm}\sum_{l\geq\max
(1,j-i)}(j-i-2l)\eta_{i+l}\mu_{j-l} - \sum_{l\geq
1}l\big[\mu_{i+l-1}\eta_{j-l-1} - \eta_{i+l-1}\mu_{j-l-1}\big] \ ,\non \\
A^{\mu \eta}_{ij} &=& \frac{j(N-i)}{N}\mu_{i}\eta_{j} +
\hspace{-4mm}\sum_{l\geq\max
(1,j-i)}(j-i-2l)\mu_{i+l}\eta_{j-l} + \sum_{l\geq
1}l\big[\mu_{i+l-1}\eta_{j-l-1} - \eta_{i+l-1}\mu_{j-l-1}\big] \ ,\non \\
A^{\mu\mu}_{ij} &=& \frac{j(N-i)}{N}\mu_{i}\mu_{j} +
\hspace{-4mm}\sum_{l\geq\max
(1,j-i)}(j-i-2l)\mu_{i+l}\mu_{j-l} + 2\sum_{l\geq1}l\mu_{i+l-1}\mu_{j-l+1} \
,\non \\
B^{\eta}_{i} &=& \frac{i(N-i)}{N}\eta_{i} \ ,\non \\
B^{\mu}_{i} &=& \frac{i(N-i)}{N}\mu_{i}\,.
\eea
As expected, the coefficients $A^{\eta\eta}_{ij}$ and $B^{\eta}_{i}$ coincide
with the expressions found in \cite{Ruhl:1995} for the bosonic Sutherland
model.
Furthermore, the coefficient functions $A_{ij}$ and $B_{i}$ are second
and first order polynomials in $\eta_{i}$ and $\mu_{i}$, respectively.
Similarly to what happens in the bosonic case the Laplace operator (\ref{e3.8})
possesses infinitely-many polynomial eigenfunctions \cite{Ruhl:1995}.
These have the form of supersymmetric analogies of the Bethe-ansatz wave
functions.

After omitting the center-of-mass motion
associated with $\si_{N}$ and $\ze_{N}$, the final expression for the
supersymmetric many-body
Sutherland Hamiltonian $h_{\rm sSuth}$ takes the form (cf. Section 3,
eq.(\ref{e2.12}))
\bea
\label{e3.10}
h^{\rm (rel)}_{\rm sSuth} &=&  \sum_{i,j=1}^{N-1}\bigg[
A^{\eta\eta}_{ij} \frac{\pa^{2}}{\pa \eta_{i}\pa\eta_{j}} +
A^{\eta\mu}_{ij} \frac{\pa^{2}}{\pa \eta_{i}\pa\mu_{j}} +
A^{\mu\eta}_{ij}\frac{\pa^{2}}{\pa \eta_{j}\pa\mu_{i}} +
A^{\mu\mu}_{ij}\frac{\pa^{2}}{\pa\mu_{j}\pa\mu_{i}}\bigg] \non \\  &+&
(\frac{1}{N} + \nu)\sum_{i=1}^{N-1}i(N-i)
\bigg[\eta_{i}\frac{\pa}{\pa \eta_{i}} + \mu_{i}\frac{\pa}{\pa\mu_{i}}\bigg]\,.
\label{hrel2}
\eea
where  $A^{\eta\eta}_{ij}, A^{\eta\mu}_{ij}, A^{\mu \eta}_{ij},
A^{\mu\mu}_{ij},
B^{\eta}_{i}, B^{\mu}_{i}$
are given in (\ref{e3.9}).
This expression can be called the {\it algebraic} form of the supersymmetric
many-body Sutherland Hamiltonian.

In superspace coordinates the Hamiltonian $h^{\rm (rel)}_{\rm sSuth}$
describing the relative motion can be written
\be
\label{e3.11}
 -h^{\rm (rel)}_{\rm sSuth} = \int d\al d\bar{\al}\sum_{i,j=1}^{N-1}
{\cal A}_{ij}\frac{\de}{\de \bar{\psi}_{j}}\frac{\de}{\de \psi_{i}} + \int d\al
{\cal W} + \int d\bar{\al}\bar{{\cal W}} \ ,
\ee
where
\be
\label{e3.12}
{\cal W} = \frac{1}{2}(\frac{1}{N} +
\nu)\sum_{i=1}^{N-1}i(N-i)\psi_{i}\frac{\de}{\de \psi_{i}} \ ,\
\bar{{\cal W}}={\cal W} (\bar{\psi}) \ ,
\ee
\be
\psi_i = \eta_{i} + \al\mu_{i}\ ,\
\frac{\de}{\de \psi_{i}} = \frac{\pa}{\pa\mu_{i}} +
\al\frac{\pa}{\pa \eta_{i}}\ ,
\ee
(cf.(\ref{psi-coord})), while
\be
\label{e3.13}
{\cal A}_{ij} = \frac{j(N-i)}{N}\psi_{i}\bar{\psi}_{j} - \sum_{l\geq\max
(1,j-i)}(j-i-2l)\psi_{i+l}\bar{\psi}_{j-l} + \sum_{l\geq
1}l\big[\bar{\psi}_{i+l-1}\psi_{j-l+1} - \psi_{i+l-1}\bar{\psi}_{j-l+1}\big]\ ,
\ee
(cf. (\ref{Aij})). The method of the calculation of the coefficients
${\cal A}_{ij}$ is similar to that presented in App.C.

It turns out that the Hamiltonian governing the relative motion can be
rewritten in
terms of the generators of a Borel subalgebra of the Lie superalgebra
$gl(N|N-1)$ (see Appendix A.2) after substituting $\ka_{i} \rar \mu_{i-1}$
and $\ta_{i} \rar \eta_{i-1}$. The final result is given in Appendix B, eq.
(\ref{e3.14}).
Since the expression (\ref{e3.14}) contains no positive-root generators
$T^{+}_i, Q^+_{\ga}$, then in accordance with the general definition
\cite{Turbiner:1994} we conclude that the supersymmetric many-body
Sutherland model (\ref{e3.1}) is exactly solvable.
The existence of the representation (\ref{e3.14}) proves that there are
infinitely-many eigenfunctions of the operator (4.3) having the form of
polynomials in the variables $(\eta_{k},\mu_{k})$.
It also leads to the conclusion that eigenfunctions of the supersymmetric
many-body Sutherland model (\ref{e3.1}) have a factorizable form being
the product of the ground-state eigenfunction multiplied by a polynomial
in the variables $(\eta_{k},\mu_{k})$. These polynomials are related to
finite-dimensional irreducible representations of the algebra $gl(N|N-1)$
in the realization (A.3).

To conclude, the supersymmetric many-body Sutherland models possess the
algebraic form (\ref{e3.10}) and also the Lie-algebraic form (\ref{e3.14})
represented by second-order polynomials in the generators of the
of the algebra $gl(N|N-1)$ with certain coefficients.

\setcounter{equation}{0}
\section{The $BC_N, B_{N}, C_{N}$ and $D_{N}$ Calogero models and their
supersymmetric extensions}

It is well-known that there is a deep connection between
the integrable Calogero-Sutherland systems and Lie algebras (see discussion
in the Introduction). This connection
is explicitly realized in the Hamiltonian reduction method
\cite{Olshanetskii:1977,Kazhdan:1978} (see also
\cite{Olshanetskii:1981,Olshanetskii:1983,Gorskii:1994}).
In particular, the celebrated many-body Calogero and Sutherland models
discussed in Section 2 are related to the Lie algebras $A_{N-1}$
and are therefore known as the Calogero and Sutherland systems of
$A_{N-1}$-type. A natural question emerges: what are the integrable systems
corresponding to the other simple Lie algebras like
$B_{N}$, $C_{N}$, $D_{N}$ etc. appearing in the Hamiltonian reduction
method. The answer and a complete classification of these systems is given
in \cite{Olshanetskii:1977} (for a review, see
\cite{Olshanetskii:1981,Olshanetskii:1983}). In the present paper we focus on
the quantum Calogero
(rational) and Sutherland (trigonometric) systems of the
$BC_{N}, B_{N}, C_{N}$ and $D_{N}$ types leaving for a future publication
the case of the exceptional algebras.

\subsection{The bosonic case}

\label{cBCD}

Unlike the $A_{N-1}$ many-body Calogero model, the
$BC_{N}, B_{N}, C_{N}$ and $D_{N}$ Calogero models are not translation
invariant, and describe systems
with boundaries. However, they are still permutation
invariant with respect to the interchange of any pair of coordinates.
The configuration space of these models is $\{x_{i}|x_{i} > 0; j<k: x_{j} <
x_{k}\}$.
In this section we will show that like the $A_{N-1}$ many-body Calogero system,
all the $BC_{N}, B_{N}, C_{N}, D_{N}$ quantum Calogero systems are not only
completely integrable, but also exactly-solvable possessing a hidden
algebra $gl(N+1)$.

It is well known that the Hamiltonians of the $BC_{N}, B_{N}, C_{N}$
Calogero models are all given by (see \cite{Olshanetskii:1977})
\be
\label{e5.1}
{\cal H}^{(c)}_{BCD} = \frac{1}{2}\sum_{i=1}^{N}[-\pa_{i}^{2} +
\om^{2}x_{i}^{2}] + g\sum_{i<j}\left[ \frac{1}{(x_{i}-x_{j})^{2}} +
\frac{1}{(x_{i}+x_{j})^{2}} \right] +
\frac{g_{2}}{2}\sum_{i=1}^{N}\frac{1}{x_{i}^{2}}
\ee
where $g = \nu(\nu - 1)$ and $g_{2} = \nu_{2}(\nu_{2} - 1)$.
When the coupling constant $g_{2}$ tends to zero, the Hamiltonian
${\cal H}^{(c)}_{BCD}$ degenerates to the Hamiltonian of the $D_{N}$
Calogero model. The ground state eigenfunction of the Hamiltonian
(\ref{e5.1}) is given by
\be
\label{e5.2}
\Psi_{0} = \left[
\prod_{i<j}|x_{i}-x_{j}|^{\nu}|x_{i}+x_{j}|^{\nu}
\prod_{i=1}^{N}|x_{i}|^{\nu_{2}} \right]
e^{-\frac{\om}{2}\sum_{i=1}^{N}x_{i}^{2}}\ ,
\ee
(cf.(\ref{e1.2})).

One should stress the point\footnote{See also the discussion in footnote 7.}
that to fixed values of the coupling constants $g, g_2$ there corresponds two
different values of the parameter
$\nu (\nu_2)$: $\nu (\nu_2) =\al (\al_2)$ and $(1-\al (\al_2))$ giving rise to
four families of eigenfunctions. Of course, $\al (\al_2)$
should be chosen in such a way as to minimize the eigenvalue
and then (\ref{e5.2}) corresponds to the ground state. The ground state can be
denoted as
$(\al, \al_2)$. The other values $\nu (\nu_2)$ inserted in (\ref{e5.2})
 describes the ground states (if normalizable) of three {\it other}
families:
$(1-\al, \al_2), (\al, 1-\al_2), (1-\al, 1-\al_2)$. If $g_2=0$, the sectors
$(\al, 1-\al_2), (1-\al, 1-\al_2)$
correspond to totally antisymmetric states with respect to the
reflections $x_i \leftarrow -x_i$. In what follows, for the sake of simplicity,
we call (\ref{e5.2}) the ground state, of course keeping
in mind the above discussion.

Following the same approach as in Sections 2-4, we make a gauge rotation
of (\ref{e5.1}) with the ground-state eigenfunction as the gauge factor;
$h^{(c)}_{BCD} = -2\Psi_{0}^{-1}{\cal H}\Psi_{0}$. A straightforward
calculation gives (we omit an additive constant)
\be
\label{e5.3}
 h^{(c)}_{BCD} = \sum_{i=1}^{N} \pa_{i}^{2} +
2\nu\sum_{i<j}\bigg[\frac{1}{x_{i}-x_{j}}(\pa_{i} - \pa_{j}) +
\frac{1}{x_{i}+x_{j}}(\pa_{i} + \pa_{j})\bigg] +
\nu_{2}\sum_{i=1}^{N}\frac{1}{x_{i}}\pa_{i} - 2\om\sum_{i=1}^{N}x_{i}\pa_{i}\,.
\ee
We call this operator the {\it rational} form of the $BC_{N}, B_{N}, C_{N}$ and
$D_N$
Calogero systems.

The Hamiltonian (\ref{e5.1}) as well as the operator (\ref{e5.3}) is
invariant under the permutation of any pair of the coordinates
$x_{i} \leftrightarrow x_{j}$ and also under the action of the reflection group
${\bf Z}_2^{\oplus N}$: $x_{i} \rar -x_{i},\ i=1,2,\ldots,N$.
An infinite set of eigenfunctions of (\ref{e5.3}) are given as polynomials
in the $x$'s and are classified by the
representations of the reflection group ${\bf Z}_2^{\oplus N}$.
For the sake of simplicity, we consider in what follows only the
 eigenfunctions of the Hamiltonian (\ref{e5.3}), which are {\it totally
symmetric} under the reflection and permutation group actions. The
{\it totally (anti)symmetric-(anti)symmetric} eigenfunctions under the
reflection and/or permutation group actions are reproduced by a change of
parameters $\nu, \nu_2$ in (\ref{e5.1}). This fact implies that it is
sufficient to look for eigenfunctions to the operator (\ref{e5.3}) of the
form
\be
\label{e5.4}
\Psi^{\nu,\nu_2}= F(x_1^2, x_2^2, \ldots, x_N^2) \ ,
\ee

Now let us construct variables which encode the permutation invariance of the
system. Analogously to what was done for the $A_{N-1}$ Calogero model
\cite{Ruhl:1995} we use as new variables the elementary
symmetric polynomials $\si_{k}$ (see (2.12)) but with  $x_i^2$ as arguments:
\be
\label{e5.6}
\sum_{k=0}^{N} \tilde\si_{k}(x_{i}^2)t^{k} = \exp\left[
\sum_{n=1}^{\infty}\frac{(-1)^{n+1}}{n}s_{2n}(x_{i})t^{n}\right]\,,
\ee
(cf. (2.10), (\ref{e2.5})). Finally, in the new variables the Hamiltonian
$h^{(c)}_{BCD}$ becomes
\be
\label{e5.7}
h^{(c)}_{BCD} = \sum_{i,j = 1}^{N} A_{ij}\frac{\pa}{\pa \tilde\si_{i}}
\frac{\pa}{\pa \tilde\si_{j}} + \sum_{j=1}^{N} B_{j}\frac{\pa}{\pa
\tilde\si_{j}}
\ee
where (cf.(3.28))
\bea
\label{e5.8}
A_{ij} &=& 4 \sum_{l\geq 0} (2l + 1 + j - i)\tilde\si_{i-l-1}\tilde\si_{j+l}
\non \\
B_{j} &=& 2[ 1 + \nu_{2} + 2\nu (N-j)](N - j + 1)
\tilde\si_{j-1} - 4\om j\tilde\si_{j} \ ,
\eea
and $\tilde{\si}_k = 0$, when $k>N$ or $k<0$. The method of the calculation
of the coefficients ${A}_{ij}, B_j$ is similar to that presented in App.C.
This expression can be called the {\it algebraic} form of the
$BC_{N}, B_{N}, C_{N}$ and $D_{N}$ Calogero Hamiltonians.

Similarly to what happened in all previously discussed cases the coefficients
$A_{ij}, B_{j}$ are polynomials of second and first degree in $\tilde\si_{k}$,
respectively. Hence, there exists a Lie algebraic form of the Hamiltonian
and $h^{(c)}_{BCD}$ can be written in terms of generators of a Borel
subalgebra of $gl(N+1)$ (see Appendix A.1) as in Section 2. The result
is presented in Appendix B, eq.(B.5).
Then in accordance with the general definition given in \cite{Turbiner:1994}
we can conclude that all $BC_{N}, B_{N}$, $C_{N}, D_{N}$ Calogero models
(\ref{e5.1}) are exactly solvable. The existence of the representation
(\ref{e5.7}) proves that there are infinitely-many eigenfunctions of the
operator (5.3) having the form of polynomials in the variables $\tilde\si_{k}$.
It also implies that totally  (anti)symmetric-(anti)symmetric eigenfunctions
with respect to  permutations and reflections of the $BC_{N}, B_{N}, C_{N}$
and $D_{N}$ Calogero models (\ref{e5.1}) have a factorizable form being the
product of the ground-state eigenfunction (5.2) multiplied by a polynomial in
the variables $\tilde\si_{k}$. These polynomials are related to
finite-dimensional irreducible representations of the Lie algebra $gl(N+1)$ in
the realization (A.1) and can be called the {\it $BC_N$ Calogero} polynomials.

\subsection{The supersymmetric extensions}

The $BC_{N}, B_{N}, C_{N}$ and $D_{N}$ Calogero models discussed above have
also natural supersymmetric extensions. These can be constructed in a
straightforward way. Let us consider the supercharges (\ref{q2.1}) with the
superpotential $W$ given by
\be
\label{e5.10}
 W = \nu\sum_{i<j} (\log|x_{i} - x_{j}| + \log|x_{i} + x_{j}|) +
\nu_{2}\sum_{i=1}^{N} \log|x_{i}|  -\frac{\om}{2}\sum_{i=1}^{N}x_{i}^{2}\,.
\ee
(cf.(\ref{q2.2})). After some  calculations the supersymmetric
Hamiltonian ${\cal H}^{(c)}_{{\rm s}BCD} = \frac{1}{2}\{Q,Q^{\dagger}\}$
of the supersymmetric $BC_{N}, B_{N}, C_{N}, D_{N}$
Calogero models emerges:
\bea
\label{e5.11}
 {\cal H}^{(c)}_{{\rm s}BCD}&=& \frac{1}{2}\sum_{i=1}^{N}
\bigg[-\frac{\pa^{2}}{\pa x_{i}^{2}} + \om^{2}x_{i}^{2}\bigg] +
\frac{\nu_{2}}{2}\sum_{i=1}^{N}\frac{1}{x_{i}^{2}}\bigg[\nu_{2} - 1 +
\tha_{i}\frac{\pa}{\pa\tha_{i}}\bigg] + \om \sum_{i=1}^{N}
\tha_{i}\frac{\pa}{\pa\tha_{i}}  \non \\ && +
\;\nu\sum_{i<j}\frac{1}{(x_{i}-x_{j})^{2}}\bigg[\nu-1 + (\tha_{i} -
\tha_{j})(\frac{\pa}{\pa\tha_{i}} - \frac{\pa}{\pa\tha_{i}})\bigg]\non
\\ && +\;
\nu\sum_{i<j}\frac{1}{(x_{i}+x_{j})^{2}}\bigg[ (\nu -1) + (\tha_{i} +
\tha_{j})(\frac{\pa}{\pa\tha_{i}} + \frac{\pa}{\pa\tha_{i}})\bigg] + C
\eea
where the constant $C= -2N [\nu(N-1) - \nu_{2} - 1]$. Again, like for the
other supersymmetric extensions the ground-state eigenfunction remains
the same as in the bosonic case and is given by (\ref{e5.4}). Making the
gauge rotation of the Hamiltonian (\ref{e5.11}):\linebreak
$h^{(c)}_{sBCD}= -2\Psi_{0}^{-1}{\cal H}^{(c)}_{{\rm s}BCD}\Psi_{0}$, we get
\bea
\label{e5.12}
 h^{(c)}_{{\rm s}BCD} &=& \sum_{i=1}^{N} \frac{\pa^{2}}{\pa x_{i}^{2}} +
2\nu\sum_{i<j}\bigg[\frac{1}{x_{i}-x_{j}}(\frac{\pa}{\pa x_{i}} -
\frac{\pa}{\pa x_{j}}) - \frac{(\tha_{i}-\tha_{j})}{(x_{i}-x_{j})^{2}}
(\frac{\pa}{\pa\tha_{i}} -\frac{\pa}{\pa\tha_{j}})\bigg] \non \\ &+&
2\nu\sum_{i<j}\bigg[\frac{1}{x_{i}+x_{j}}(\frac{\pa}{\pa x_{i}} +
\frac{\pa}{\pa x_{j}}) - \frac{(\tha_{i}+\tha_{j})}{(x_{i}+x_{j})^{2}}
(\frac{\pa}{\pa\tha_{i}} +\frac{\pa}{\pa\tha_{j}})\bigg] \non \\ &-&
 2\om\sum_{i=1}^{N}\bigg[x_{i}\frac{\pa}{\pa x_{i}} +
\tha_{i}\frac{\pa}{\pa\tha_{i}}\bigg] +
\nu_{2}\sum_{i=1}^{N}\bigg[\frac{1}{x_{i}}\frac{\pa}{\pa x_{i}} -
\frac{\tha_{i}}{x_{i}^{2}}\frac{\pa}{\pa\tha_{i}}\bigg] \,.
\eea
We call this operator the {\it rational} form of the supersymmetric
$BC_{N}, B_{N}, C_{N}$ Calogero systems.

The superspace coordinates $\chi_{i}$ which are invariant under the
permutation and parity transformations are defined through the relation
\be
\label{e5.13}
\sum_{k=0}^{N} \chi_{k}(\phi_{i}^2)t^{k} = \exp\left[
\sum_{n=1}^{\infty}\frac{(-1)^{n+1}}{n}s_{2n}(\phi_{i})t^{n}\right]\ ,
\ee
(cf.(5.5) and (3.19)), where $\phi_i = x_i + \al \tha_i$ as in Sections 3
and 4, and
$\chi_{k}(\phi_{i}^2) = \tilde{\si}_{k}(x_i^2) + \al\tilde{\ze}_k(x_i^2,2
x_i\tha_i)$. Here $\tilde{\si}$ and $\tilde{\ze}$
are the elementary symmetric polynomials defined in (\ref{e2.4}) but of the
new arguments:  $x \rar x^2, \tha \rar 2x\tha$.
Using the same technique as in Sections 3-4, one can find the representation
of the supersymmetric models directly from the analogous representation
for the bosonic cases (for details see App.C). Finally, in the superspace
coordinates (\ref{e5.13}) the Hamiltonian $h^{(c)}_{{\rm s}BCD}$ has the form
\be
\label{e5.14}
h^{(c)}_{{\rm s}BCD} = \int d\al d\bar{\al} \sum_{i,j=1}^{N} {\cal
A}_{ij}\frac{\pa}{\pa \bar{\chi}_{i}}\frac{\pa}{\pa \chi_{j}} + \int d\al
\sum_{j=1}^{N} {\cal B}_{j}\frac{\pa}{\pa \chi_{j}}
\ee
where
\bea
{\cal A}_{ij} &=& 2 \sum_{l\geq 0} \Big\{ 2(2l + 1 + j -
i)\chi_{i-l-1}\bar{\chi}_{j+l} - l[\chi_{j+l}\bar{\chi}_{i-l-1} -
\bar{\chi}_{j+l}\chi_{i-l-1} \cr
\phantom{\Bigg\{}&+& \chi_{i+l-1}\bar{\chi}_{j-l} -
\bar{\chi}_{i+l-1}\chi_{j-l}]\Big\}  \, , \non \\
{\cal B}_{j} &=& 2[ 1 + \nu_{2} + 2\nu (N-j)](N - j + 1)\chi_{j-1} -
4\om j\chi_{j}\ . \label{e5.14,5}
\eea
For completeness we give the form of $h^{(c)}_{{\rm s}BCD}$ in components
\bea
\label{e5.15}
h^{(c)}_{{\rm s}BCD}  &=&
\sum_{i,j=1}^{N}\bigg[A^{\tilde{\si}\tilde{\si}}_{ij}
\frac{\pa^{2}}{\pa \tilde{\si}_i\pa\tilde{\si}_j} +
A^{\tilde{\si}\tilde{\ze}}_{ij}
\frac{\pa^{2}}{\pa \tilde{\si}_i\pa\tilde{\ze}_j} +
A^{\tilde{\ze}\tilde{\si}}_{ij}
\frac{\pa^{2}}{\pa \tilde{\ze}_i\pa\tilde{\si}_j} +
A^{\tilde{\ze}\tilde{\ze}}_{ij}
\frac{\pa^{2}}{\pa \tilde{\ze}_j\pa\tilde{\ze}_i}\bigg] \non \\
&\hspace{-15mm}+&\hspace{-10mm}\sum_{j=1}^{N}\bigg\{2[1+ \nu_{2}
+2\nu(N-j)](j-N+1)\bigg[\tilde{\si}_{j-1}\frac{\pa}{\pa \tilde{\si}_{j}} +
\tilde{\ze}_{j-1}\frac{\pa}{\pa \tilde{\ze}_{j}} \bigg] \non \\
&\hspace{-15mm}-&\hspace{-10mm} 4\om j\bigg[\tilde{\si}_{j}\frac{\pa}{\pa
\tilde{\si}_{j}}
+ \tilde{\ze}_{j}\frac{\pa}{\pa \tilde{\ze}_{j}}\bigg]\bigg\} \,,
\eea
where
\[
A^{\tilde{\si}\tilde{\si}}_{ij} = 4 \sum_{l\geq 0} (2l + 1 + j -
i)\tilde{\si}_{i-l-1}\tilde{\si}_{j+l}\ ,
\]
\[
A^{\tilde{\si}\tilde{\ze}}_{ij} = 2 \sum_{l\geq 0}\Big[ 2(2l + 1 + j -
i)\tilde{\si}_{i-l-1}\tilde{\ze}_{j+l} -
l[\tilde{\si}_{j-l-1}\tilde{\ze}_{i+l}-\tilde{\si}_{i+l}\tilde{\ze}_{j-l-1} +
\tilde{\si}_{j-l}\tilde{\ze}_{i+l-1}-\tilde{\si}_{i+l-1}\tilde{\ze}_{j-l}]\Big]
\,
\]
\[
A^{\tilde{\ze}\tilde{\si}}_{ij} = 2 \sum_{l\geq 0}\Big[ 2(2l + 1 + j -
i)\tilde{\ze}_{i-l-1}\tilde{\si}_{j+l} +
l[\tilde{\si}_{j-l-1}\tilde{\ze}_{i+l}-\tilde{\si}_{i+l}\tilde{\ze}_{j-l-1} +
\tilde{\si}_{j-l}\tilde{\ze}_{i+l-1}-
\tilde{\si}_{i+l-1}\tilde{\ze}_{j-l}]\Big]\,
\]
\be
\label{e5.16}
A^{\tilde{\ze}\tilde{\ze}}_{ij} = 4 \sum_{l\geq 0}\Big[ (2l + 1 + j -
i)\tilde{\ze}_{i-l-1}\tilde{\ze}_{j+l} - l\tilde{\ze}_{i+l}\tilde{\ze}_{j-l-1}
-l\tilde{\ze}_{i+l-1}\tilde{\ze}_{j-l}\Big]\,. \label{Att}
\ee
>From the expression (\ref{e5.15}) with the coefficients (\ref{e5.16}) it
follows that $h^{(c)}_{{\rm s}BCD}$ can be written as a quadratic polynomial
in the generators of a Borel subalgebra of $gl(N+1|N)$ and, hence, the $BC_{N}$
Calogero model is exactly solvable.

The existence of the representation of (\ref{e5.15}) in terms of the generators
of a Borel subalgebra of $gl(N+1|N)$ (see (B.6)), proves that the operator
$h^{(c)}_{{\rm s}BCD}$ has infinitely-many eigenfunctions having the form
of polynomials. These polynomials are related to finite-dimensional irreducible
representations of the algebra $gl(N+1|N)$ in the realization (A.3) and can
be called the supersymmetric $BC_N$ Calogero polynomials.

So the supersymmetric $BC_{N}, B_{N}, C_{N}$ and $D_{N}$ Calogero models
possess the algebraic form and also the Lie-algebraic form (B.6) being
represented by second-order polynomials in the generators of the of the
algebra $gl(N+1|N)$ with certain coefficients.

\setcounter{equation}{0}
\section{The $BC_{N}, B_{N}$, $C_{N}$ and $D_{N}$ Sutherland models and
their supersymmetric extensions}
\label{sBCD}

Similarly to what was done in Section 5 for the
$BC_{N}, B_{N}, C_{N}$ and $D_{N}$ Calogero models the present section is
devoted
to a consideration of the $BC_{N}, B_{N}, C_{N}$ and $D_{N}$
Sutherland models.

\subsection{The bosonic case}

The Hamiltonians for $BC_{N}, B_{N}, C_{N}, D_{N}$
Sutherland models are special cases of the general $BC_N$ Hamiltonian
\cite{Olshanetskii:1977}
\bea
\label{e6.1}
{\cal H}^{(s)}_{BCD} &=& -\frac{1}{2}\sum_{i=1}^{N} \frac{\pa^{2}}{\pa
x_{i}^{2}} + \frac{g}{4}\sum_{i<j}\left[
\frac{1}{\sin^{2}(\frac{1}{2}(x_{i}-x_{j}))} +
\frac{1}{\sin^{2}(\frac{1}{2}(x_{i}+x_{j}))} \right] +
\frac{g_{2}}{4}\sum_{i=1}^{N}\frac{1}{\sin^{2}(x_{i})} \non \\ &&+
\frac{g_{3}}{8}\sum_{i=1}^{N}\frac{1}{\sin^{2}(\frac{x_{i}}{2})}
\eea
where $g = \nu(\nu - 1)$, $g_{2} = \nu_{2}(\nu_{2} - 1)$ and $g_{3} =
\nu_{3}(\nu_{3} + 2\nu_{2} - 1)$.
>From the general Hamiltonian the $B_{N}$, $C_{N}$ and $D_{N}$ cases
are obtained as follows:
\begin{itemize}
\item $B_{N}$ case:  $\nu_{2}=0$,
\item $C_{N}$ case:  $\nu_{3}=0$, and
\item $D_{N}$ case:  $\nu_{2}=\nu_{3}=0$.
\end{itemize}

The ground state wave function is \cite{Olshanetskii:1983,Bernard:1995}
\be
\label{e6.2}
\Psi_{0} = \left[
\prod_{i<j}|\sin(\frac{1}{2}(x_{i}-x_{j}))|^{\nu}
|\sin(\frac{1}{2}(x_{i}+x_{j}))|^{\nu}\prod_{i=1}^{N}
|\sin(x_{i})|^{\nu_{2}}|\sin(\frac{x_{i}}{2})|^{\nu_{3}} \right] \,.
\ee

Again, one should emphasize \footnote{See also the discussion
in footnote 7 and in Section 5.1} that to a fixed value of the coupling
constant $g (g_2) [g_3]$ there corresponds two different values of the
parameter
$\nu (\nu_2)  [\nu_3] : \nu (\nu_2) [\nu_3] =\al (\al_2) [\al_3]$ and
$1-\al (1-\al_2) [1-\al_3]$ giving rise to eight families of eigenfunctions.
In order to get the ground state the parameters $\al, \al_2, \al_3$ should
be chosen in such a way as to minimize the eigenvalue and then (\ref{e6.2})
corresponds to the ground state. It can be denoted
$(\al, \al_2, \al_3)$. The other values of $\nu, \nu_2, \nu_3$, if inserted
in (\ref{e6.2}), describe (provided that the corresponding wavefunctions are
normalizable) the ground states of the remaining seven {\it other} families
of eigenstates:
\[
(\al, \al_2, 1-\al_3)\ ,\ (\al, 1-\al_2, \al_3)\ ,\ (\al, 1-\al_2, 1-\al_3)\ ,
\]
\[
(1-\al, \al_2, \al_3)\ ,\ (1-\al, 1-\al_2, \al_3)\ ,\ (1-\al, \al_2, 1-\al_3)\
,
\]
\[
(1-\al, 1-\al_2, 1-\al_3)\ .
\]
So taking different values of $\nu, \nu_2, \nu_3$ in (6.2) one can exhaust
all the types of ground states corresponding to the different families
mentioned above.\footnote{In what follows, for the sake of simplicity,
we continue to call (\ref{e6.2}) the ground state, of course, keeping
in mind above discussion.}

Using the same approach as in Sections 2-4, we make a gauge rotation
of (\ref{e6.1}) with the ground-state eigenfunction as gauge factor,
$h^{(s)}_{BCD} = -2\Psi_{0}^{-1}{\cal H}^{(s)}_{BCD}\Psi_{0}$.
A straightforward calculation leads to the operator
(we omit an additive constant)
\bea
\label{e6.3}
 h^{(s)}_{BCD} &=& \sum_{i=1}^{N} \pa_{i}^{2} +
\nu\sum_{i<j}[\cot(\frac{1}{2}(x_{i}-x_{j}))(\pa_{i} - \pa_{j}) +
\cot(\frac{1}{2}(x_{i}+x_{j}))(\pa_{i} + \pa_{j})] \non \\ &+&
\nu_{2}\sum_{i=1}^{N}\cot(x_{i})\pa_{i} +
\frac{\nu_{3}}{2}\sum_{i=1}^{N}\cot(\frac{x_{i}}{2})\pa_{i}\,.
\eea
It is worth mentioning that if the operator (\ref{e6.3}) is written in the
coordinates $e^{ix}$, it appears in its {\it rational} form \cite{Bernard:1995}
(cf. the rational form of the $A_{N-1}$ Sutherland Hamiltonian in
\cite{Sutherland:1971}).

According to the above discussion about eight different families of
eigenfunctions, it is sufficient  to study the operator
(6.3) for generic $\nu, \nu_2, \nu_3$ and consider only eigenfunctions which
are symmetric under reflections. In particular, this implies that
eigenfunctions of (6.3) have the form
\be
\label{e6.4}
P^{\nu,\nu_{2},\nu_3}= F(\cos x_1, \cos x_2, \ldots, \cos x_N) \,,
\ee
where $F$ is permutationally symmetric.

Next we encode the symmetry properties of the problem studied by introducing
the permutation and reflection-invariant periodic variables
$\hat{\si}_{k}(\cos x)$, which satisfy
\be
\label{e6.5}
\sum_{k=0}^{N} \hat\si_{k}(\cos(x_{i}))t^{k} = \exp\left[
\sum_{n=1}^{\infty}\frac{(-1)^{n+1}}{n}s_{n}(\cos(x_{i}))t^{n}\right]\,,
\ee
(cf.(4.8), (5.11)). Let us emphasize that these variables (6.5) are
characterized by the same period as the original Hamiltonian (6.1).
In these variables the Hamiltonian $h^{(s)}_{BCD}$ becomes
\be
\label{e6.6}
h^{(s)}_{BCD} = \sum_{i,j = 1}^{N} A_{ij}\frac{\pa}{\pa \hat\si_{i}}
\frac{\pa}{\pa \hat\si_{j}} + \sum_{j=1}^{N} B_{j}\frac{\pa}{\pa \hat\si_{j}}
\ee
where
\bea
\label{e6.7}
A_{ij} &=&N\hat\si_{i-1}\hat\si_{j-1} - \sum_{l\geq 0}\Big[ (i - l)
\hat\si_{i-l}\hat\si_{j+l} + (l + j - 1)\hat\si_{i-l-1}\hat\si_{j+l-1} \non \\
&&- (i - 2  - l)\hat\si_{i-2-l}\hat\si_{j+l}
- (l + j + 1)\hat\si_{i-l-1}\hat\si_{j+l+1}\Big]\non \\
B_{j} &=& \frac{\nu_{3}}{2}(j-N-1)\hat\si_{j-1} - [\nu_{2} + \frac{\nu_{3}}{2}
+ 1 + \nu(2N-j-1)]j\hat\si_{j}  \non \\&& - \nu(N -j+1)(N-j+2)\hat\si_{j-2}\ ,
\label{sAB}
\eea
and $\hat\si_{k} = 0$, when $k<0$ or $k>N$. The method of the calculation of
the coefficients ${A}_{ij}, B_j$ is similar to that presented in App.C.
This expression can be called the
{\it algebraic} form of the $BC_{N}, B_{N}, C_{N}$ and $D_{N}$ Sutherland
Hamiltonians. Similarly to what happened in all previously discussed bosonic
cases the coefficients $A_{ij}, B_{j}$ are polynomials of second and first
degree in $\hat{\si}_{k}$, respectively. Hence, the Hamiltonian
$h^{(s)}_{BCD}$ can be written in terms of generators of the Borel subalgebra
of $gl(N+1)$ realized as first order differential operators
(see Appendix A.1) as in Sections 2 and 5.1. The result is given in eq.
(B.7) (see Appendix B).

Then in accordance with the general definition given in \cite{Turbiner:1994}
we conclude that all $BC_{N}, B_{N}$, $C_{N}, D_{N}$ Sutherland models
(\ref{e6.1}) are exactly solvable. The existence of the representation
(\ref{e6.6}) proves that there are infinitely-many eigenfunctions of the
operator (5.3) having the form of polynomials in the variables $\hat\si_{k}$.
It also implies that totally (anti)symmetric-(anti)symmetric eigenfunctions
with respect to  permutations and reflections of the $BC_{N}, B_{N}, C_{N}$
and $D_{N}$ Sutherland models (\ref{e6.1}) have a factorizable form being the
product of the ground-state eigenfunction (6.2) multiplied by a polynomial in
the variables $\hat\si_{k}$. These polynomials are related to
finite-dimensional
irreducible representations of the Lie algebra $gl(N+1)$ in the realization
(A.1). They can be called the $BC_{N}$ Jack-Sutherland polynomials.

\subsection{The supersymmetric extensions}

The bosonic $BC_{N}, B_{N}$, $C_{N}$ and $D_{N}$ Sutherland models (\ref{e6.1})
have natural supersymmetric extensions. In this subsection we will construct
these models and show that, as was the case for the other supersymmetric
extensions discussed in Sections 3,4 and 5.2, these models are also
exactly-solvable.
Let us introduce the supercharges (\ref{q2.1}) with the superpotential
$W$ given by
\bea
\label{e6.8}
 W &=& \nu\sum_{i<j} \Big[\log|\sin(\frac{1}{2}(x_{i} - x_{j}))| +
\log|\sin(\frac{1}{2}(x_{i} + x_{j}))|\Big] + \nu_{2}\sum_{i=1}^{N}
\log|\sin(x_{i})| \non \\ &+& \nu_{3}\sum_{i=1}^{N}
\log|\sin(\frac{x_{i}}{2})|\,.
\eea
(cf.(3.4),(5.8)). From the general formula
${\cal H}^{(s)}_{{\rm s}BCD} = \frac{1}{2}\{Q,Q^{\dagger}\}$, we derive
the supersymmetric Hamiltonians of the $BC_{N}, B_{N}$, $C_{N}$ and $D_{N}$
Sutherland models:
\bea
\label{e6.9}
 {\cal H}^{(s)}_{{\rm s}BCD}&=& -\frac{1}{2}\sum_{i=1}^{N} \frac{\pa^{2}}{\pa
x_{i}^{2}} +  \frac{\nu}{4}\sum_{i<j}\Big[
\frac{1}{\sin^{2}(\frac{1}{2}(x_{i}-x_{j}))}[\nu - 1 +(\tha_{k} -
\tha_{l})(\frac{\pa}{\pa\tha_{k}} -
\frac{\pa}{\pa\tha_{l}})] \non \\ &+&
\frac{1}{\sin^{2}(\frac{1}{2}(x_{i}+x_{j}))}[\nu - 1 + (\tha_{k} +
\tha_{l})(\frac{\pa}{\pa\tha_{k}} +
\frac{\pa}{\pa\tha_{l}})] \Big] + \sum_{i=1}^{N}\Bigg\{\frac{\nu_{3}}{8}\big[
\frac{1}{\sin^{2}(\frac{x_{i}}{2})}[\nu_{3} - 1 \cr &+&
\tha_{i}\frac{\pa}{\pa\tha_{i}}]\Big] +\frac{\nu_{2}}{2}\Big[
\frac{1}{\sin^{2}(x_{i})}[\nu_{2} +
2\nu_{3} - 1 + \tha_{i}\frac{\pa}{\pa\tha_{i}}] \Big]\Bigg\} + C \,,
\eea
where $C=-\frac{1}{2}\sum_{i=1}^{N}[\nu(N-i)+\nu_{2}+\frac{\nu_{3}}{2}]^{2}$.
As should be familiar by now, the next step is to introduce the gauge-rotated
Hamiltonian
$h^{(s)}_{{\rm s}BCD} = -2\Psi_{0}^{-1}{\cal H}^{(s)}_{{\rm s}BCD}\Psi_{0}$.
After some calculations we get
\bea
\label{e6.10}
 h^{(s)}_{BCD} &=& \sum_{i=1}^{N} \pa_{i}^{2} +
\nu\sum_{i<j}\bigg[\cot(\frac{1}{2}(x_{i}-x_{j}))(\frac{\pa}{\pa x_{i}} -
\frac{\pa}{\pa x_{j}}) - \frac{1}{2}\frac{(\tha_{i}-
\tha_{j})}{\sin^{2}(\frac{1}{2}(x_{i}-x_{j}))}(\frac{\pa}{\pa\tha_{i}}
-\frac{\pa}{\pa\tha_{j}}) \bigg] \non \\ &+&
\nu\sum_{i<j}\bigg[\cot(\frac{1}{2}(x_{i}+x_{j}))(\frac{\pa}{\pa x_{i}} +
\frac{\pa}{\pa x_{j}}) - \frac{1}{2}\frac{(\tha_{i}+
\tha_{j})}{\sin^{2}(\frac{1}{2}(x_{i}+x_{j}))}(\frac{\pa}{\pa\tha_{i}} +
\frac{\pa}{\pa\tha_{j}})\bigg] \non \\&+&
\nu_{2}\sum_{i=1}^{N}\bigg[\cot(x_{i})\frac{\pa}{\pa x_{i}} -
\frac{\tha_{i}}{\sin^{2}(x_{i})}\frac{\pa}{\pa\tha_{i}}\bigg] +
\frac{\nu_{3}}{2}\sum_{i=1}^{N}\bigg[\cot(\frac{x_{i}}{2})\frac{\pa}{\pa
x_{i}}-
\frac{1}{2}\frac{\tha_{i}}{\sin^{2}(\frac{x_{i}}{2})}\frac{\pa}{\pa\tha_{i}}
\bigg]\ .
\eea
It is worth mentioning that if the operator (\ref{e6.10}) is written in the
coordinates $e^{ix}$ it appears in its {\it rational} form
\footnote{This expression should be compared to the rational form of the
supersymmetric $A_{N}$ Sutherland model (4.3) and the $BC_N$ Sutherland model
(6.3).}.

As in Sections 3, 4 and 5.2 it is convenient to use the superspace formalism
and to
introduce the superspace coordinates $\chi_{i}$, which satisfies:
\be
\label{e6.11}
\sum_{k=0}^{N} \chi_{k}(\cos(\phi_{i}))t^{k} = \exp\left[
\sum_{n=1}^{\infty}\frac{(-1)^{n+1}}{n}s_{n}(\cos(\phi_{i}))t^{n}\right]\,.
\ee
Here $\chi_{k}(\cos(\phi_{i})) = \hat{\si}_{k}(\cos x_i) + \al
\hat{\ze}_k(\cos x_i,-\tha_i\sin x_i)$, where $\hat{\si}_{k}$ and
$\hat{\ze}_k$ are the elementary symmetric polynomials defined in (\ref{e2.4})
but of new arguments: $x \rar \cos x, \tha \rar - \tha \sin x$.
In terms of these variables the Hamiltonian (6.10) becomes
\be
\label{e6.12}
h^{(s)}_{{\rm s}BCD} = \int d\al d\bar{\al} \sum_{i,j=1}^{N} {\cal
A}_{ij}\frac{\pa}{\pa \bar{\chi}_{i}}\frac{\pa}{\pa \chi_{j}} + \int d\al
\sum_{j=1}^{N} {\cal B}_{j}\frac{\pa}{\pa \chi_{j}}\,,
\ee
where
\bea
{\cal A}_{ij} &=&N\chi_{i-1}\bar{\chi}_{j-1} - \sum_{l\geq 0}\Big[ (i -
l)\chi_{i-l}\bar{\chi}_{j+l} + (l + j - 1)\chi_{i-l-1}\bar{\chi}_{j+l-1} \non
\\ &&- (i - 2  - l)\chi_{i-2-l}\bar{\chi}_{j+l} - (l + j +
1)\chi_{i-l-1}\bar{\chi}_{j+l+1} + l[\chi_{i+l-1}\bar{\chi}_{j-l-1} -
\bar{\chi}_{i+l-1}\bar{\chi}_{j-l-1} \non \\&+&  \chi_{i+l+1}\chi_{j-l-3}-
\chi_{i+l+1}\bar{\chi}_{j-l-3} +  \chi_{i+l+1}\bar{\chi}_{j-l-1} -
\bar{\chi}_{i+l+1}\chi_{j-l-1} ]\Big]\, ,\non \\
{\cal B}_{j} &=& \frac{\nu_{3}}{2}(j-N-1)\chi_{j-1} - (\nu_{2} + \frac{\nu_{3}}{2} + 1
+
\nu(2N-j-1))j\chi_{j}  \non \\&& - \nu(N -j+1)(N-j+2)\chi_{j-2} \,.
\eea
The method of the calculation of the coefficients
${\cal A}_{ij}, {\cal B}_j$ is similar to that presented in App.C.
In components $h^{(s)}_{{\rm s}BCD}$ becomes
\bea
\label{e6.14}
h^{(s)}_{{\rm s}BCD}  &=&  \sum_{i,j=1}^{N}\bigg[A^{\hat{\si}\hat{\si}}_{ij}
\frac{\pa^{2}}{\pa \hat{\si}_i\pa\hat{\si}_j} + A^{\hat{\si}\hat{\ze}}_{ij}
\frac{\pa^{2}}{\pa \hat{\si}_i\pa\hat{\ze}_j} + A^{\hat{\ze}\hat{\si}}_{ij}
\frac{\pa^{2}}{\pa \hat{\ze}_i\pa\hat{\si}_j} + A^{\hat{\ze}\hat{\ze}}_{ij}
\frac{\pa^{2}}{\pa \hat{\ze}_j\pa\hat{\ze}_i}\bigg] \non \\
&\hspace{-15mm}+&\hspace{-10mm}\sum_{j=1}^{N}\bigg\{
\frac{\nu_{3}}{2}(j-N-1)\bigg[\hat{\si}_{j-1}\frac{\pa}{\pa \hat{\si}_{j}} +
\hat{\ze}_{j-1}\frac{\pa}{\pa \hat{\ze}_{j}}\bigg] - (\nu_{2} +
\frac{\nu_{3}}{2} + 1 + \nu(2N-j-1))j\cr
 &&\hspace{-5mm}\cdot\bigg[\hat{\si}_{j}\frac{\pa}{\pa
\hat{\si}_{j}} +
\hat{\ze}_{j}\frac{\pa}{\pa \hat{\ze}_{j}}\bigg] - \nu(N
-j+1)(N-j+2)\bigg[\hat{\si}_{j-2}\frac{\pa}{\pa \hat{\si}_{j}} +
\hat{\ze}_{j-2}\frac{\pa}{\pa \hat{\ze}_{j}}\bigg]\bigg\} \,,
\eea
where
\[
A^{\hat{\si}\hat{\si}}_{ij} = N\hat{\si}_{i-1}\hat{\si}_{j-1} -
\sum_{l\geq 0}\Big[ (i - l)\hat{\si}_{i-l}\hat{\si}_{j+l} + (l + j -
1)\hat{\si}_{i-l-1}\hat{\si}_{j+l-1}
\]
\[
 - \;(i - 2  - l)\hat{\si}_{i-2-l}\hat{\si}_{j+l} - (l + j +
1)\hat{\si}_{i-l-1}\hat{\si}_{j+l+1}\Big]
\]
\[
A^{\hat{\si}\hat{\ze}}_{ij} = N\hat{\si}_{i-1}\hat{\ze}_{j-1} -
\sum_{l\geq 0}\Big[ (i - l)\hat{\si}_{i-l}\hat{\ze}_{j+l} + (l + j -
1)\hat{\si}_{i-l-1}\hat{\ze}_{j+l-1}
\]
\[
 -\; (i - 2  - l)\hat{\si}_{i-2-l}\hat{\ze}_{j+l} - (l + j +
1)\hat{\si}_{i-l-1}\hat{\ze}_{j+l+1}
\]
\[
-\;l[\hat{\si}_{j-l-1}(\hat{\ze}_{i+l-1}+\hat{\ze}_{i+l+1})+
\hat{\si}_{i+l+1}(\hat{\ze}_{j-l-3}-\hat{\ze}_{j-l-1})]\Big]
\]
\[
A^{\hat{\ze}\hat{\si}}_{ij} = N\hat{\ze}_{i-1}\hat{\si}_{j-1} -
\sum_{l\geq 0}\Big[ (i - l)\hat{\ze}_{i-l}\hat{\si}_{j+l} + (l + j -
1)\hat{\ze}_{i-l-1}\hat{\si}_{j+l-1}
\]
\[
 -\; (i - 2  - l)\hat{\ze}_{i-2-l}\hat{\si}_{j+l} - (l + j +
1)\hat{\ze}_{i-l-1}\hat{\si}_{j+l+1}
\]
\[
+\; l[\hat{\si}_{j-l-1}(\hat{\ze}_{i+l-1}+\hat{\ze}_{i+l+1})
+\hat{\si}_{i+l+1}(\hat{\ze}_{j-l-3}-\hat{\ze}_{j-l-1}) ]\Big]
\]
\bea
\label{e6.15}
A^{\hat{\ze}\hat{\ze}}_{ij} &=& N\hat{\ze}_{i-1}\hat{\ze}_{j-1} -
\sum_{l\geq 0}\Big[ (i -
l)\hat{\ze}_{i-l}\hat{\ze}_{j+l} + (l + j -
1)\hat{\ze}_{i-l-1}\hat{\ze}_{j+l-1} \cr &-& (i - 2  -
l)\hat{\ze}_{i-2-l}\hat{\ze}_{j+l} - (l + j +
1)\hat{\ze}_{i-l-1}\hat{\ze}_{j+l+1} \cr
\phantom{\Bigg\{}&+&
l[\hat{\ze}_{i+l-1}\hat{\ze}_{j-l-1}-\hat{\ze}_{i+l+1}\hat{\ze}_{j-l-3}
+2\hat{\ze}_{i+l+1}\hat{\ze}_{j-l-1}]\Big]\, .
\eea

>From the expression (\ref{e6.14}) with the coefficients (\ref{e6.15})
it follows, that $h^{(s)}_{{\rm s}BCD}$ can be written as a quadratic
polynomial in the generators of a Borel subalgebra of $gl(N+1|N)$ and,
hence, the $BC_{N}$ Sutherland model is exactly solvable.

The existence of the representation of (\ref{e6.14}) in terms of the generators
of a Borel subalgebra of $gl(N+1|N)$ (see (B.8)) proves that there are
infinitely-many eigenfunctions of (\ref{e6.14}) having the form of polynomials.
These polynomials are related to finite-dimensional irreducible representations
of the algebra $gl(N+1|N)$ in the realization (A.3). These polynomials can
be called the supersymmetric $BC_N$ Sutherland polynomials.

So the supersymmetric $BC_{N}, B_{N}, C_{N}$ and $D_{N}$ Sutherland models
possess an algebraic form and also the Lie-algebraic form (B.8) being
represented by second-order polynomials in the generators of the of the
algebra $gl(N+1|N)$ with certain coefficients.

\setcounter{equation}{0}
\section{Conclusion}

In this paper we have described the rational, algebraic and
Lie-algebraic forms of  the integrable $A_N-B_N-C_N-D_N$ rational (Calogero)
and trigonometric (Sutherland) Hamiltonians in addition to the superspace
expressions for the supersymmetric generalizations of the these models. We have
shown that
\begin{quote}
{\it All Hamiltonians of the integrable $A_N-B_N-C_N-D_N$ rational (Calogero)
and trigonometric (Sutherland) models possess the {\bf same} hidden algebra
$gl(N+1)$ and can be represented by second-degree polynomials in the generators
of a Borel subalgebra of the $gl(N+1)$-algebra. If the configuration space
is parametrized by permutationally symmetric coordinates $v$, the
Hamiltonians have a triangular form and preserve the flag of spaces of
inhomogeneous polynomials
\[
\label{e7.1}
{\cal P}_n \ =\ \mbox{\rm span} \{ v_1^{n_1} v_2^{n_2} v_3^{n_3}
\ldots v_{N}^{n_{N}} |0 \leq \sum n_i \leq n \}\ ,
\]
Consequently, each Hamiltonian possesses one or several infinite families
of polynomial eigenfunctions.

All Hamiltonians of the supersymmetric generalizations of the
$A_N-B_N-C_N-D_N$ rational (Calogero) and trigonometric (Sutherland) models
possess the {\bf same} hidden algebra $gl(N+1|N)$ and can be represented by
second-degree polynomials in the generators of a Borel subalgebra of the
$gl(N+1|N)$-algebra. If the configuration space is parametrized by
permutationally symmetric coordinates $v, \ka$
\footnote{where $\ka^2=0$.},
the Hamiltonians have a triangular form and preserve the flag of spaces of
inhomogeneous polynomials
\[
\label{e7.2}
{\cal P}_n \ =\ \mbox{\rm span} \{ v_1^{n_1} v_2^{n_2}
\ldots v_{N}^{n_{N}} \ka_1^{m_1} \ka_2^{m_2}
\ldots \ka_{N}^{m_{N}} |0 \leq \sum n_i+\sum m_i \leq n , m_i=0,1 \}\ ,
\]
Consequently, each Hamiltonian has one or several infinite
families of polynomial eigenfunctions. The integrability of the $A_N$
supersymmetric system was proven in \cite{Shastry:1993}; as for the
$B_N-C_N-D_N$ systems, it is an open question.}
\end{quote}

As an interesting issue for future study we would like to mention the question
about whether there exists
a Lie algebraic description of the higher $A_N-B_N-C_N-D_N$ rational
(Calogero) and trigonometric (Sutherland) Hamiltonians. It is very probable
that this will be the case and, perhaps, even the Lax operator can be
represented in terms of $gl(N)$-generators. There are almost no doubts
that it will be possible to extend the analysis of Sections 2-6 to
the case of the exceptional simple Lie algebras. In particular, a
special case of the one-parametric $G(2)$ rational and trigonometric
Hamiltonians corresponding to three-body interactions only
\cite{Wolfes:1974,Quesne:1996} possesses a hidden $gl(3)$-algebra and,
hence, is expressible in terms of the $gl(3)$ generators \cite{Quesne:1996}
\footnote{Recently, a Lie algebraic form for the general $G(2)$ rational and
trigonometric models was found \cite{Capella:1997}.}.

The question about the integrability of the above supersymmetric models leads
naturally to the question about the possibility to develop the Hamiltonian
reduction method for the supersymmetric cases and attempt to connect these
models to 2$d$ supersymmetric gauge theories along the lines of
\cite{Gorskii:1994}.

An alternative description of the supersymmetric models discussed in this
paper can be obtained by using the matrix representation of the $\tha^{\al}$'s
in terms of the Pauli matrices, $\si^{\pm,3}$:
\bea
\tha^{1} &=& \si^{-}\otimes 1\otimes 1 \otimes \cdots \non \\
\frac{\pa}{\pa \tha^{1}} &=& \si^{+}\otimes 1\otimes 1 \otimes \cdots \non \\
\tha^{2} &=& \si^{3} \otimes \si^{-}\otimes 1 \otimes \cdots \non \\
\frac{\pa}{\pa \tha^{2}} &=& \si^{3}\otimes\si^{+}\otimes 1 \otimes \cdots
\non \\
 & \vdots&
\eea
In this formalism our supersymmetric models become matrix
generalizations of the `scalar' $A_N-B_N-C_N-D_N$ Calogero and Sutherland
models. For the $A_N$ case our matrix models are particular cases of the
general
construction with internal degrees of freedom proposed in \cite{Minahan:1993}.

The existence of the Lie algebraic form for the exactly-solvable
$A_N-B_N-C_N-D_N$ Calogero and Sutherland models can be considered as a
good starting point to investigate whether there exist other
exactly-solvable problems
and/or quasi-exactly-solvable generalizations (for definitions see,
for instance, \cite{Turbiner:1988,Morozov:1990}). The first step
in applying the Lie algebraic formalism to the search for
quasi-exactly-solvable problems was taken in \cite{Minzoni:1996}.

As a final comment we would like to mention that a great challenge
in the subject is the search for solutions of the $N$-body elliptic model
(for a description see, for example, \cite{Olshanetskii:1983}). For the
two-body case, the rational and algebraic forms have been known since Hermite's
days. Recently, it was found that the 2-body case (the Lame operator) admits
a Lie algebraic form with the hidden algebra $gl(2)$ and is a
{\it quasi-exactly-solvable} problem \cite{Turbiner:1989} (see also
\cite{Turbiner:1994}). It gives a hint that the three-body (and more
generally the $N$-body) elliptic
problem has to be quasi-exactly-solvable if any analytic solution exists
whatsoever. However,
all attempts
to find a rational, algebraic or Lie algebraic form even for the three-body
elliptic model have failed so far.

\section*{Acknowledgements}

A.T. thanks the Institute for Theoretical Physics, Chalmers University of
Technology, where this work was initiated, for its kind hospitality.
A.T. is grateful to the NORDITA for financial support during his visit to
G\"oteborg.

\newpage
\section*{Appendices}

\appendix
\setcounter{equation}{0}

\section{Representations of the Lie algebra $gl(N)$ and the Lie superalgebra
$gl(N|M)$.}

\subsection{$gl(N)$}

The generators of the Lie algebra $gl(N)$ can be realized by first order
differential operators. For our purposes we need one of the simplest
realizations of $gl(N)$ acting on the real (complex) space of dimension
$(N-1)$.
This is the space of minimal dimension where this algebra can act.
The generators can be represented in the following form:
\bea \label{a1}
E_{0i} &=& J_i^- = { \partial \over \partial \tau_i}  \ ,   \quad i=2,3,\ldots
, N \ ,\non \\
E_{ij} &=& J_{i,j}^0 = \tau_i J_j^-=\tau_i { \partial \over \partial \tau_j}\
,\quad i,j=2,3,
\ldots , N \ , \non \\
E_{00} &=& J^0 = n - \sum_{k=2}^N \tau_k \frac{\partial}{\partial \tau_k} \ ,
\non \\
E_{i0} &=& J_i^+ = \tau_i J^0\ , \quad i=2,3,\ldots , N \,,
\eea
where the parameter $n \in {\bf R}$ $({\bf C})$. One of the generators,
namely $J^{0} + \sum_{p=2}^{N} J_{p,p}^{0}$, is proportional
to a constant and, if it is removed, we end up with the algebra
$sl(N)$. The generators $J_{i,j}^0$ form the algebra of the vector
fields of $sl(N-1)$, which is a subalgebra of $gl(N)$.
If $n$ is a non-negative integer, the representation (\ref{a1})
becomes the finite-dimensional representation acting on the space
of polynomials in $(N-1)$ variables of the following type
\be
\label{a2}
V_n(t)\ =\ \mbox{span} \{ \tau_2^{n_2} \tau_3^{n_3} \tau_4^{n_4} \ldots
\tau_{N}^{n_{N}} | 0 \leq \sum n_i \leq n\}\ .
\ee
This representation corresponds to a Young tableau with one row and $n$ blocks
and is irreducible. The positive-root generators $J_i^+$ define the
highest-weight vector. If the $J_{i}^{+}$'s are removed, the remaining
generators form a
Borel subalgebra.

\subsection{$gl(N|M)$}

Similarly to what was done for the algebra $gl(N)$ one can construct a
representation of the Lie superalgebra $gl(N|M)$ in terms of first order
differential operators over the direct sum of a even space and an odd
space. Again the simplest realization of $gl(N|M)$ act on the space
$C^{(N-1|M)}$ spanned by the the even variables $(\tau^i| i=1,2,\ldots,N-1))$
and the
odd variables $(\ka^{\ga}| \ga=1,2,\ldots, M)$ and is given by
\bea
\label{b1}
E_{ij} = J^0_{ij} = \tau^{i}\frac{\pa}{\pa \tau^{j}}\,, && E_{00} = J^{0} = - n
+
\sum_{i}\tau^{j}\frac{\pa}{\pa \tau^{j}} + \sum_{\ga}\ka^{\ga}\frac{\pa}{\pa
\ka^{\ga}} \non \\
E_{i0} = J^{+}_{i} = \tau^{i}J^{0}\,, && E_{0i} = J_{i}^{-} = \frac{\pa}{\pa
\tau^{i}}
\non \\
E_{\ga i} = Q_{\ga i} = \ka^{\ga}\frac{\pa}{\pa \tau^{i}}\,, && E_{i\ga} =
\bar{Q}_{i \ga} = \tau^{i}\frac{\pa}{\pa \ka^{\ga}} \non \\ E_{0\ga} =
Q^{-}_{\ga}
= \frac{\pa}{\pa \ka^{\ga}}\,, && E_{\ga 0} = Q^{+}_{\ga} = \ka^{\ga}T^{0}
\non\\
E_{\ga\beta} = T^{0}_{\ga\beta} &=& \ka^{\ga}\frac{\pa}{\pa \ka^{\beta}}
\label{gen}
\eea
where the parameter $n \in {\bf R}$ $({\bf C})$.
Here the indices run over the following values, $i=1\ldots N-1$ and
$\ga=1\ldots M$. The subset $\{E_{ij},E_{\ga i},E_{i \ga},E_{\ga\beta}\}$
form the ``vector field'' representation of $gl(N-1|M)$. The algebra has a
${\bf Z}_{2}$ gradation and it is clear that the elements with an odd
number of Greek indices are odd and the elements with an even number are even.
Using the composite index notation $I=(0,i,\ga)$, the commutation relations of
the superalgebra can be written (cf. the commutation relations in the defining
$(N+M)\times(N+M)$ supermatrix representation, with
$(E_{IJ})_{MN} = \de_{IN}\de_{JM}$)
\be
\ba{lcccl}
{}[ E_{IJ},E_{KL} ] &=& \de_{JK}E_{IL} - \de_{IL}E_{KJ}\,, && ({\rm
both\;even}) \non \\
{}[ E_{IJ},E_{KL} ] &=& \de_{JK}E_{IL} - \de_{IL}E_{KJ}\,, && (E_{IJ}\; {\rm
even},\;E_{KL}\;{\rm odd}) \non \\
{}\{ E_{IJ},E_{KL}\} &=& \de_{JK}E_{IL} + \de_{IL}E_{KJ}\,, && ({\rm
both\;odd})
\ea \,.\label{comm}
\ee

The Lie superalgebra $gl(N|M)$ is not simple. There are two different
cases: (i) $N\neq M$ and (ii) $N = M$. Let us consider the first case,
when $N\neq M$. Removing the unit element $\sum_{I=0}^{N+M}E_{II}$
we are left with the superalgebra $sl(N|M)$, which is simple.
The ``diagonal'' generators for this case are given by
\be
H_{II} = \left\{ \ba{lcr} E_{00} - E_{11}\,, && I = 0 \\
E_{II}-E_{I+1,I+1}\,, && 1\leq I \leq N-1 \\
E_{II}+E_{I+1,I+1}\,, && I = N \\
E_{II}-E_{I+1,I+1}\,, && N+1\leq I \leq N+M-1
\ea \right.
\ee

When $N=M$ the situation is slightly more complicated. There are two
one dimensional abelian ideals, which must be removed in order to make
the algebra simple.

The generators of $gl(N|M)$ defined in (\ref{gen}) act in the space
spanned by the monomials $\{\tau_{1}^{i_{1}}\cdots
\tau_{N-1}^{i_{N-1}}\ka_{1}^{\de_{1}}\cdots\ka_{M}^{\de_{M}}\}$.
Here $\de_{k}$ equals either $0$ or $1$. When $n$ (in the expression
for $T^{0}$ above) is an integer we have finite-dimensional highest
weight representations whose highest weight vectors satisfy
$\sum_{k=1}^{N-1} i_{k} + \sum_{k=1}^{M} \de_{k} = n$. These
representations form a flag of spaces.

\setcounter{equation}{0}
\section{The Lie algebraic forms of the $A_{N-1}, BC_{N}, B_{N}$, $C_{N}$ and
$D_{N}$ Hamiltonians}
In this Appendix we collect all Lie algebraic forms of the
Hamiltonians for the (supersymmetric) $A_{N-1}, BC_{N}, B_{N}$, $C_{N}$ and
$D_{N}$ Calogero and Sutherland models found in \cite{Ruhl:1995} and in
the present paper.

\begin{itemize}

\item  {\it $A_{N-1}$ Calogero model}

The Lie algebraic form of the many-body Calogero  model is the following
\cite{Ruhl:1995}
\begin{eqnarray}
\label{e12}
h_{\rm Cal} &=& \sum_{j=2}^{N} \left\{ \frac{(N-j+1)(j-1)}{N} (J_{j-1,j})^{2} -
2 \sum_{\ell =1}^{j-1} \ell J_{j+\ell -1,j} J_{j-\ell -1,j} \right\}
\nonumber \\ &\hspace{-5mm}+&\hspace{-5mm}
2\hspace{-2mm} \sum_{2 \leq k < j \leq N} \left\{ \frac{(N-j+1)(k-1)}{N}
J_{j-1,j}
J_{k-1,k}
- \sum_{\ell =1}^{k-1} (j-k+2\ell ) J_{j+\ell -1,j} J_{k-\ell -1,k}
\right\} \cr &-& 2\om \sum_{k=2}^{N} k J_{k,k} -
\left(\frac{1}{N} + \nu \right) \sum_{k=2}^{N}(N-k+2)(N-k+1) J_{k-2,k}
\end{eqnarray}
where the following notations are introduced for the $gl(N)$ generators (A.1):
$J_{ij} \equiv J^{0}_{ij}$, $J_{0,k}^{0} \equiv J_{k}^{-}$, $J_{1,k}^{0}
\equiv 0$.

The supersymmetric Calogero model has the Lie algebraic form:
\bea
\label{e2.16}
h_{\rm sCal}^{\rm (rel)} &=& \sum_{i,j=2}^{N}\Big\{ \frac{(N-i+1)(j-1)}
{N}[J_{i-1,j}J_{j-1,i} +
J_{i-1,i}T_{j-1,j} + J_{j-1,j}T_{i-1,i} \cr &-& T_{i-1,j}T_{j-1,i}]
+\sum_{l>\max
(1,j-i)}(j-i-2l)[J_{i+l-1,j}J_{j-l-1,i} + J_{i+l-1,i}T_{j-l-1,j} \cr &+&
J_{j-l-1,j}T_{i+l-1,i} - T_{i+l-1,j}T_{j-l-1,i} ] \Big\} + \sum_{l\geq
1}l[T_{j-2-l,j}J_{i+l,i} - T_{i+l,j}J_{j-2-l,i} \cr &+& T_{i+l,i}J_{j-2-l,j} -
T_{j-2-l,i}J_{i+l,j} -2T_{i+l,j}T_{j-2-l,i}] -
2\om \sum_{j=2}^{N}j[J_{jj} + T_{jj}] \cr &-&
(\frac{1}{N} + \nu)\sum_{j=2}^{N}(N-j+2)(N-j+1)[J_{j-2,j} + T_{j-2,j}] \,,
\eea
where the following conventions are used for the $gl(N|N-1)$ generators
(A.3):
$J_{ij} \equiv J^{0}_{ij} = 0$ and $T_{ij} \equiv T^{0}_{ij} 0$, if $i$ and/or
$j$ is less than 2 or
greater than $N$.

\item {\it $A_{N-1}$ Sutherland model  }

The Lie algebraic form of the many-body Sutherland model in terms of
the $gl(N)$ generators is the following
\cite{Ruhl:1995}
\begin{eqnarray} \label{bossuth}
-h_{\rm Suth} &=& \sum_{j=1}^{N-1} \left\{ \frac{j(N-j)}{N} (J_{j,j})^{2} -
2 \sum_{l=1}^{j} l J_{j+l,j} J_{j-l,j} \right\} + \nu \sum_{l=1}^{N-1} l (N-l)
J_{l,l} \nonumber \\
&+& 2\hspace{-2mm} \sum_{1 \leq k < j \leq N-1}
\left\{ \frac{k(N-j)}{N} J_{j,j} J_{k,k} -
\sum_{l=1}^{k} (j-k+2l) J_{j+l,j} J_{k-l,k} \right\} \,.
\end{eqnarray}
Here we have used the identifications $J_{ij} \equiv J^{0}_{ij}$, $J_{0,i}
\equiv J^{-}_{i}\equiv J_{N,i}$, $J_{1,i} = 0$.
The supersymmetric many-body Sutherland model written in terms of
the $gl(N|N-1)$ generators has the form
\bea
\label{e3.14}
-h^{\rm (rel)}_{\rm sSuth} &=& \sum_{i,j=1}^{N-1}\Big\{
\frac{j(N-i)}{N}[J_{i,j}J_{j,i} + J_{i,i}T_{j,j} + J_{j,j}T_{i,i} -
T_{i,j}T_{j,i}] \non \\  &\hspace{-6mm}+&\hspace{-6mm}\sum_{l>\max
(1,j-i)}(j-i-2l)[J_{i+l,j}J_{j-l,i} +
J_{i+l,i}T_{j-l,j} + J_{j-l,j}T_{i+l,i} - T_{i+l,j}T_{j-l,i} ] \Big\} \non \\
&+&\sum_{l\geq 1}l[T_{j-l-1,j}J_{i+l-1,i} - T_{i+l-1,j}J_{j-l-1,i} +
T_{i+l-1,i}J_{j-l-1,j}- T_{j-l-1,i}J_{i+l-1,j} \cr &-& 2T_{i+l-1,j}T_{j-l+1,i}]
+\sum_{i=1}^{N-1}i(N-i)[\nu J_{i,i} + (\frac{2}{N} + \nu)T_{i,i}]\,.
\eea
In (\ref{e3.14}) $J_{ij} \equiv j^{0}_{ij} = 0$ and $T_{i,j} \equiv T^{0}_{ij}
=0$ are by definition
zero when $i$ and/or $j$ is greater than $N-1$ or less than $1$.

\item  {\it $BC_N$, $B_N$, $C_N$ and $D_N$ Calogero models}

The bosonic $BC_N$, $B_N$, $C_N$ and $D_N$ Calogero models have the following
Lie algebraic form in terms of the $gl(N+1)$ generators
\bea
\label{e5.9}
 h^{(c)}_{BCD} &=&  4 \sum_{i,j=1}^{N}\sum_{l\geq 0} (2l + 1 + j -
i)J_{i-l-1,j}J_{j+l,i}- 4\om\sum_{i=1}^{N}iJ_{i,i} \cr
&\hspace{-3.5mm}+&\hspace{-2.5mm}
2\sum_{i=1}^{N}\left\{ [ 1 + \nu_{2} + 2\nu (N-j)](N - j + 1) -
\frac{N}{2}(N+3-2i)\right\} J_{i-1,i}\,,
\eea
and the supersymmetric extensions have the form
\bea
  h^{(c)}_{{\rm s}BCD} &=&  2 \sum_{i,j=1}^{N}\sum_{l\geq 0} \Big[ 2(2l + 1 + j
-
i)[J_{i-l-1,j}J_{j+l,i} + J_{i-l-1,j}T_{j+l,i} + T_{i-l-1,j}J_{j+l,i} \cr
\phantom{\Bigg\{} &-& T_{i-l-1,j}T_{j+l,i}] + l[T_{j-l-1,j}J_{i+l,i} -
T_{i+l,j}J_{j-l-1,i} - T_{i+l-1,j}J_{j-l,i} \cr
\phantom{\Bigg\{} &+& T_{j-l,j}J_{i+l-1,i} - T_{j-l-1,i}J_{i+l,j} +
T_{i+l,i}J_{j-l-1,j} +
T_{i+l-1,i}J_{j-l,j} - T_{j-l,i}J_{i+l-1,j} \cr
&+& T_{i+l,j}T_{j-l-1,i} + T_{i+l-1,j}T_{j-l,i}]\Big] + \sum_{i=1}^{N}2[ 1 +
\nu_{2} + 2\nu (N-j)](N -
j + 1) \cr
&&\hspace{-5mm} \cdot\Big\{ \frac{N}{2}(N+3-2i)T_{i-1,i} -
\frac{N}{2}(N+3-2i)J_{i-1,i}\Big\} - 4\om\sum_{i=1}^{N}i[J_{i,i} + T_{i,i}]\,.
\eea
in terms of the $gl(N+1|N)$ generators. In the above two formulas, as well as
in the two formulas below, we have used that $J_{ij} \equiv J^{0}_{ij}=0$ and
$T_{ij}\equiv T^{0}_{ij} = 0$ when $i$ and/or $j$ is greater than $N$ or less
than $1$.

\item  {\it $BC_N$, $B_N$, $C_N$ and $D_N$ Sutherland models }

Finally, the Lie algebraic form of the bosonic $BC_N$, $B_N$, $C_N$ and $D_N$
Sutherland models is
\bea
h^{(s)}_{BCD} &=& \sum_{i,j=1}^{N}\Bigg[ NJ_{i-1,j}J_{j-1,i} - \sum_{l\geq
0}\Big[ (i - l)J_{i-l,j}J_{j+l,i} +
(l + j - 1)J_{i-l-1,j}J_{j+l-1,i} \cr &-& (i - 2  - l)J_{i-2-l,j}J_{j+l,i}
- (l + j + 1)J_{i-l-1,j}J_{j+l+1,i}\Big] \cr &+&
\sum_{j=1}^{N}\Bigg[\frac{\nu_{3}}{2}(j-N-1)J_{j-1,j}
- (\nu_{2} + \frac{\nu_{3}}{2} + 1 + N
+ \nu(2N-j-1))jJ_{j,j} \cr &-& [\nu(N -j+1)(N-j+2) + 2N]J_{j-2,j}\Bigg]
\eea
in terms of the $gl(N+1)$ generators, and the Lie algebraic form of their
supersymmetric extensions is
\bea
h^{(s)}_{{\rm s}BCD} &=& \sum_{i,j=1}^{N}\Bigg\{ N(J_{i-1,j}J_{j-1,i} -
T_{i-1,j}T_{j-1,i}) - \sum_{l\geq 0}\Big[ (i - l)[J_{i-l,j}J_{j+l,i} +
T_{i-l,j}J_{j+l,i} \cr
\phantom{\Bigg[}  &\hspace{-8mm}+&\hspace{-5mm}
J_{i-l,j}T_{j+l,i} -
T_{i-l,j}T_{j+l,i}] +
(l + j - 1)[J_{i-l-1,j}J_{j+l-1,i} + J_{i-l-1,j}T_{j+l-1,i} \cr
\phantom{\Bigg[} &\hspace{-8mm}+&\hspace{-5mm} T_{i-l-1,j}J_{j+l-1,i} -
T_{i-l-1,j}T_{j+l-1,i}] - (i - 2
- l)[J_{i-2-l,j}J_{j+l,i} + T_{i-2-l,j}J_{j+l,i} \cr\phantom{\Bigg[}
\phantom{\Bigg[}&\hspace{-8mm}+&\hspace{-5mm}
J_{i-2-l,j}T_{j+l,i} - T_{i-2-l,j}T_{j+l,i}]
- (l + j + 1)[J_{i-l-1,j}J_{j+l+1,i} + T_{i-l-1,j}J_{j+l+1,i} \cr
\phantom{\Bigg[}&\hspace{-8mm}+&\hspace{-5mm}
J_{i-l-1,j}T_{j+l+1,i} - T_{i-l-1,j}T_{j+l+1,i} ] + l[J_{j-l-1,j}(T_{i+l-1,i} +
T_{i+l+1,i}) \cr
\phantom{\Bigg[}&\hspace{-8mm}+&\hspace{-5mm}  J_{i+l+1,j}(T_{j-l-3,i} -
T_{j-l-1,i})- J_{j-l-1,i}(T_{i+l-1,j} + T_{i+l+1,j})
\cr&\hspace{-8mm}-&\hspace{-5mm}   J_{i+l+1,i}(T_{j-l-3,j} - T_{j-l-1,j}) -
T_{I+l+1,j}(T_{j-l-1,i} - T_{j-l-3,i}) \cr
\cr
&\hspace{-8mm}-&\hspace{-5mm} (T_{i+l-1,j} + T_{i+l-1,j})T_{j-l-1,i}]\Big]
\Bigg\} +
\sum_{j=1}^{N}\Bigg[\frac{\nu_{3}}{2}(j-N-1)[J_{j-1,j} + T_{j-1,j}] \cr
&\hspace{-8mm}-&\hspace{-5mm}
(\nu_{2} + \frac{\nu_{3}}{2} + 1
+ \nu(2N-j-1))j[J_{j,j} + T_{j,j}] - Nj[J_{j,j} - T_{j,j}] \cr
&\hspace{-8mm}-&\hspace{-5mm}[\nu(N
-j+1)(N-j+2)](J_{j-2,j} + T_{j-2,j}) - 2N(J_{j-2,j} - T_{j-2,j})\Bigg]
\eea
in terms of the $gl(N+1|N)$ generators.

\end{itemize}

In conclusion, let us emphasize the fact that all Lie algebraic
forms of the supersymmetric models (B.2), (B.4), (B.6), (B.8) contain
no fermionic (odd) generators.

\setcounter{equation}{0}
\section{Deriving the Lie algebraic forms: technical details}
\label{appC}

In this appendix we will give some details of the derivation of the Lie
algebraic forms of the Hamiltonians discussed in this paper. We will exemplify
the method for the $BC_N$ Calogero models. The other cases are tackled in a
similar way.

\noindent
{\bf(A)\ \it Bosonic $BC_N$ Calogero model}.\\
Our final goal is to derive
(5.6)--(5.7). Let us recall that the gauge-rotated $BC_N$ Hamiltonian is given
by
\be
  h^{(c)}_{BCD} = \sum_{i=1}^{N} \pa_{i}^{2} +
2\nu\sum_{i<j}\bigg[\frac{1}{x_{i}-x_{j}}(\pa_{i} - \pa_{j}) +
\frac{1}{x_{i}+x_{j}}(\pa_{i} + \pa_{j})\bigg] +
\nu_{2}\sum_{i=1}^{N}\frac{1}{x_{i}}\pa_{i} - 2\om\sum_{i=1}^{N}x_{i}\pa_{i}\,.
\ee
After the change of variables $x_{i} \rar \tilde{\si}_{i}$ the Hamiltonian
becomes
\bea
h^{(c)}_{BCD} &=& \sum_{\ell, k=1}^{N} \sum_{i=1}^{N}
\frac{\pa\tilde{\si}_{\ell}}{\pa
x_{i}}\frac{\pa\tilde{\si}_{k}}{\pa
x_{i}}\frac{\pa}{\pa\tilde{\si}_{k}}\frac{\pa}{\pa\tilde{\si}_{l}}  +
\sum_{k=1}^{N} \sum_{i=1}^{N}\frac{\pa^{2}\tilde{\si}_{k}}{\pa
x_{i}^{2}}\frac{\pa}{\pa\tilde{\si}_{k}}  -
2\om\sum_{k=1}^{N}\sum_{i=1}^{N}x_{i}\frac{\pa \tilde{\si}_{k}}{\pa
x_{i}}\frac{\pa}{\pa\tilde{\si}_{k}} \cr &+&
\nu\sum_{k=1}^{N}\sum_{i\neq j}\Big[\frac{1}{x_{i}+x_{j}}(\frac{\pa
\tilde{\si}_{k}}{\pa x_{i}} + \frac{\pa\tilde{\si}_{k}}{\pa x_{j}} ) +
\frac{1}{x_{i}-x_{j}}(\frac{\pa \tilde{\si}_{k}}{\pa x_{i}} -
\frac{\pa\tilde{\si}_{k}}{\pa x_{j}} )\Big]\frac{\pa}{\pa\tilde{\si}_{k}} \cr
&+& \nu_{2}\sum_{k=1}^{N}\sum_{i=1}^{N}\frac{1}{x_{i}}\frac{\pa
\tilde{\si}_{k}}{\pa x_{i}}\frac{\pa}{\pa\tilde{\si}_{k}} \label{c2}
\,.
\eea
As an example, let us express
\be
      \tilde {  B}_k =  \sum_{i\neq j}^{N}\Big[\frac{1}{x_{i}+x_{j}}
(\frac{\pa \tilde{\si}_{k}}{\pa
x_{i}} + \frac{\pa\tilde{\si}_{k}}{\pa x_{j}} ) +
\frac{1}{x_{i}-x_{j}}(\frac{\pa \tilde{\si}_{k}}{\pa x_{i}} -
\frac{\pa\tilde{\si}_{k}}{\pa x_{j}} )\Big] \label{term1}
\ee
and
\be
\tilde {  A}_{\ell,k} = \sum_{i=1}^{N}\frac{\pa\tilde{\si}_{\ell}}{\pa
x_{i}}\frac{\pa\tilde{\si}_{k}}{\pa x_{i}} \label{term2}
\ee
in terms of the $\tilde{\si}$ variables. The approach we will follow
is to represent the generating functions of (C.3) and (C.4)
\[
\tilde {  A}(t,p)=
\sum_{k=0}^{N} \tilde {  A}_{\ell,k} t^{\ell} p^{k}\quad ,\quad
\tilde {  B}(t)= \sum_{k=0}^{N} \tilde {  B}_k t^{k} \ ,
\]
in terms of the generating function of the $\tilde{\si}$ variables
\be
G(t)=
\sum_{k=0}^{N} \tilde{\si}_{k}(x)t^{k} = \exp
[\sum_{n=1}^{\infty}\frac{(-1)^{n+1}}{n}s_{2n}(x)t^{n} ]\ . \label{c5}
\ee

Let us begin by calculating the generating function $\tilde {  B}(t)$
for (\ref{term1})
\bea
&& \tilde {  B}(t)\ =\ \sum_{i\neq j}\bigg[\frac{1}{x_{i}-x_{j}}(\frac{\pa}
{\pa x_{i}} - \frac{\pa}{\pa x_{j}}) +
\frac{1}{x_{i}+x_{j}}(\frac{\pa}{\pa x_{i}} + \frac{\pa}{\pa x_{j}})\bigg]G(t)
\cr
&=& 2\sum_{i\neq j}\bigg\{ \sum_{m=1}^{\infty} (-1)^{m+1}
\bigg[\frac{x_{i}^{2m-1}
+ x_{j}^{2m-1}}{x_{i} + x_{j}} + \frac{x_{i}^{2m-1} - x_{j}^{2m-1}}{x_{i} -
x_{j}}\bigg]t^{m}\bigg\}G(t) \ .
\eea
Using the relation
\be
\frac{x_{i}^{2n-1} + x_{j}^{2n-1}}{x_{i} + x_{j}} +
\frac{x_{i}^{2n-1} - x_{j}^{2n-1}}{x_{i} - x_{j}} =
2\sum_{l=0}^{n-1}x_{i}^{2l}x_{j}^{2(n-1)-2l}\,,
\ee
and making rather obvious mathematical transformations,
one can show that
\be
\tilde {  B}(t)\ =\
\Big[4t^3\frac{\pa^2}{\pa t^2} - 4(2N-2)t^2\frac{\pa}{\pa t} +
4tN(N-1)\Big]G(t)\,.
\ee
Substituting (\ref{c5}) in (C.8) we find that $\tilde{B}_{k}$ in
(\ref{term1}) is
\be
\tilde {  B}_k\ =\  4(N-k+1)(N-k)\tilde{\si}_{k-1}\ .
\ee

Next we proceed to the calculation of (\ref{term2}). The corresponding
generating function can be written as
\be
\tilde {  A}(t,p)\ =\ \sum_{i=1}^{N}\frac{\pa G(p)}{\pa x_{i}}
\frac{\pa G(t)}{\pa x_{i}} =
4\sum_{i=1}^N \sum_{k,\ell=1}^{\infty} (-1)^{k+\ell} x_i^{2(k+\ell)-2}
t^k p^{\ell} G(t)G(p)\,.
\ee
Changing the summation variables to $r=k+\ell$ and $u=k-\ell$ one obtains
that
\be
\tilde {  A}(t,p) = 4\sum_{i=1}^N \sum_{r=2}^{\infty}
(-1)^{r}s_{2(r-1)}(tp)^{\frac{r}{2}}
\sum_u
\bigg(\frac{t}{p}\bigg)^{\frac{u}{2}} G(t)G(p)\,. \label{c11}
\ee
The sum over $u$ is easily determined
\be
\sum _u \bigg(\frac{t}{p}\bigg)^{\frac{u}{2}} =
\bigg(\frac{t}{p}\bigg)^{-\frac{r-2}{2}}\left(
\frac{1 - \bigg(\frac{t}{p}\bigg)^{r-1}}{1 -\frac{t}{p}}\right)
\label{sum1}\,.
\ee
Using this formula it is easy to transform (\ref{c11}) to the following
expression for the generating function:
\be
\tilde {  A}(t,p)\ =\
\frac{4t}{1-\frac{t}{p}}\Big[t\frac{\pa}{\pa t} - p\frac{\pa}{\pa p}\Big]
G(t)G(p)\,. \label{c13}
\ee
Expanding the rhs of (C.13) we arrive at the final expression for
(\ref{term2})
\be
\tilde {  A}_{\ell,k}\ =\
4\sum_{q\geq 0}(2q + 1 +l - k)\tilde{\si}_{k-q-1}\tilde{\si}_{\ell+q}\,.
\ee
The other terms in (\ref{c2}) can be calculated using the same method
which finally leads to the results presented in (\ref{e5.7}), (\ref{e5.8}).

\noindent
{\bf(B)\ \it Supersymmetric $BC_N$ Calogero model}. \\
For the supersymmetric models the calculations are similar to those carried
out for the bosonic
models. In terms of the superspace $\phi$-coordinates (3.11) the gauge
rotated Hamiltonian (\ref{e5.12}) is written as
\bea
  h^{(c)}_{sBCD} &=& \int d\al d\bar{\al} \sum_{i=1}^{N} \frac{\de}{\de
\phi_{i}(\bar{\al})}\frac{\de}{\de \phi_{i}(\al)}  + 2\nu\int d\al
\sum_{i<j}\bigg[\frac{1}{\phi_{i}(\al)-\phi_{j}(\al)}(\frac{\de}{\de
\phi_{i}(\al)} -\frac{\de}{\de \phi_{j}(\al)}) \cr & +&
\frac{1}{\phi_{i}(\al)+\phi_{j}(\al)}(\frac{\de}{\de \phi_{i}(\al)} +
\frac{\de}{\de \phi_{j}(\al)})\bigg] + \nu_{2}\int d\al
\sum_{i=1}^{N}\frac{1}{\phi_{i}(\al)}\frac{\de}{\de \phi_{i}(\al)} \cr
&-&2\om\int d\al \sum_{i=1}^{N}\phi_{i}(\al)\frac{\de}{\de \phi_{i}(\al)}\,.
\eea

Now we make the change of variables
$\phi_{i} \rar \chi_{i}$ (see (5.11)) using the relation (\ref{e3.20}).
Finally, an expression similar to (\ref{c2}) emerges except for the fact
that the bosonic coordinates are replaced by supercoordinates with
integration over the Grassmann variables:
\bea
h^{(c)}_{sBCD} &=& \sum_{i=1}^N \int d\al d\bar{\al}d\beta d\bar{\beta}
\frac{\de \chi_{k}
(\bar{\beta})}{\de \phi_{i}(\bar{\al})}\frac{\de \chi_{l}(\beta)}{\de
\phi_{i}(\al)}\frac{\de}{\de \chi_{l}(\bar{\beta})}\frac{\de}{\de
\chi_{l}(\beta)} \cr &-&2\nu\int d\al d\beta
\sum_{i<j}\sum_{k=1}^{N}\bigg[\frac{1}{\phi_{i}(\al)-\phi_{j}(\al)}
(\frac{\de \chi_{k}(\beta)}{\de
\phi_{i}(\al)} + \frac{\de \chi_{k}(\beta)}{\de \phi_{j}(\al)})
\cr  && \hspace{30mm} +\;\;
\frac{1}{\phi_{i}(\al)+\phi_{j}(\al)}(\frac{\de\chi_{k}(\beta)}{\de
\phi_{i}(\al)} +
\frac{\de \chi_{k}(\beta)}{\de \phi_{j}(\al)})\bigg]\frac{\de}{\de
\chi_{k}(\beta)} \cr &\hspace{-40mm}-&\hspace{-22mm} \nu_{2}\int d\al d\beta
\sum_{i=1}^{N}\sum_{k=1}^{N}\frac{1}{\phi_{i}(\al)}\frac{\de\chi_{k}
(\beta)}{\de \phi_{i}(\al)}\frac{\de}{\de \chi_{k}(\beta)} + 2\om\int
d\al \sum_{i=1}^{N}\sum_{k=1}^{N}\phi_{i}(\al)\frac{\de\chi_{k}(\beta)}
{\de \phi_{i}(\al)}\frac{\de}{\de \chi_{k}(\beta)}\,.
\eea

For terms linear in
derivatives the process of calculation is exactly the same as for the
bosonic case. The final result can be obtained from the bosonic calculation
simply by replacing the $\si$-coordinates by $\chi$-supercoordinates.
However, for the term quadratic in derivatives:
\be
\sum_{i=1}^N \int d\al d\bar{\al}d\beta d\bar{\beta} \frac{\de \chi_{k}
(\bar{\beta})}{\de \phi_{i}(\bar{\al})}\frac{\de \chi_{l}(\beta)}{\de
\phi_{i}(\al)}\frac{\de}{\de \chi_{l}(\bar{\beta})}\frac{\de}{\de
\chi_{l}(\beta)} \equiv \int d\beta d\bar{\beta}
\tilde{\cal A}^{(s)}_{k,\ell}\frac{\de}{\de\chi_{l}(\bar{\beta})}
\frac{\de}{\de\chi_{l}(\beta)}\ , \label{c16}
\ee
a slight complication arises. Again our strategy is to make use of a generating
function of the $\chi$ coordinates
\be
{\cal G}(\beta,p)=\sum_{k=0}^{N} \chi_{k}(\phi_{i}(\beta))p^{k} =
\exp \bigg[\sum_{n=1}^{\infty}\frac{(-1)^{n+1}}{n}s_{2n}(\phi_{i}(\beta))
p^{n} \bigg]\,.
\label{c17}
\ee
in order to represent the generating function for
$\tilde{\cal A}^{(s)}_{k,\ell} =
\sum_{i=1}^N \int d\al d\bar{\al} \frac{\de \chi_{k}
(\bar{\beta})}{\de \phi_{i}(\bar{\al})}\frac{\de \chi_{\ell}(\beta)}{\de
\phi_{i}(\al)}$:
\be
\tilde{\cal A}^{(s)}(t,p) =
\sum_{k,\ell=1}^{N}\tilde{\cal A}^{(s)}_{k,\ell}t^k p^{\ell}\,.
\ee
In what follows we use the notation
$\phi_i := \phi_i (\beta)$ and $\bar{\phi}_i := \phi_i (\bar{\beta})$.

Let us first note that
\be
\tilde{\cal A}^{(s)}(t,p) =
4\sum_{i=1}^N \sum_{m,n=1}^{\infty} (-1)^{n+m} \bar{\phi}_i^{2n-1}
\phi_i^{2m-1} t^n p^m {\cal G}(\beta,p){\cal G}(\bar{\beta},t)\ ,
\label{a}
\ee
(cf.(C.10)).
After some simple mathematical transformations similar to those leading to
(C.11) $\tilde{\cal A}^{(s)}(t,p)$ can be rewritten as
\bea
&& 2 \sum_{r=2}^{\infty} (-1)^{r}(tp)^{\frac{r}{2}}\sum_u [s_{2r-2}
(\bar{\phi}) +
s_{2r-2}(\phi)]\bigg(\frac{t}{p}\bigg)^{\frac{u}{2}}{\cal G}(\beta,p)
{\cal G}(\bar{\beta},t) \cr &+&
4\sum_{r=2}^{\infty} (-1)^{r}(tp)^{\frac{r}{2}}\sum_u u[\frac{s_{2r-2}
(\bar{\phi}) - s_{2r-2}(\phi)}{2(r
-1)}]\bigg(\frac{t}{p}\bigg)^{\frac{u}{2}}{\cal G}(\beta,p){\cal G}
(\bar{\beta},t)\,.  \label{c.20}
\eea
Using (\ref{sum1}) together with the relation
\be
\sum _u u\bigg(\frac{t}{p}\bigg)^{\frac{u}{2}} =
\bigg(\frac{t}{p}\bigg)^{-\frac{r-2}{2}}
\frac{1}{(1 - \frac{t}{p})^2}\left[ -(r-2) + r
\frac{t}{p} - r\bigg(\frac{t}{p}\bigg)^{r-1} + (r-2)
\bigg(\frac{t}{p}\bigg)^{r}\right]\,,
\ee
(\ref{c.20}) takes the form
\bea
&&4 \frac{t}{1-\frac{t}{p}}\sum_{r=1}^{\infty} (-1)^{r+1}[s_{2r}(\phi)p^r -
s_{2r}(\bar{\phi})t^r]{\cal G}(\beta,p){\cal G}(\bar{\beta},t)
\cr &-& 2 \frac{t}{p}\frac{t+p}{(1-\frac{t}{p})^2}
\sum_{r=1}^{\infty} \frac{(-1)^{r+1}}{r}[s_{2r}(\bar{\phi}) - s_{2r}(\phi)]
(t^r - p^r){\cal G}(\beta,p){\cal G}(\bar{\beta},t)\ . \label{b}
\eea
Evidently, the first term in (\ref{b}) is equal to
\be
 \frac{4t}{1-\frac{t}{p}}
[t\frac{\pa}{\pa t} -
p\frac{\pa}{\pa p}]{\cal G}(\beta,p){\cal G}(\bar{\beta},t)
\ee
(cf.(\ref{c13})). In order to calculate the second term in (\ref{b})
we need to use two relations:
\be
-\sum_{r=1}^{\infty} \frac{(-1)^{r+1}}{r}[s_{2r}(\bar{\phi}) - s_{2r}
(\phi)](t^r - p^r) = \left[ e^{-\sum_{r=1}^{\infty} \frac{(-1)^{r+1}}
{r}[s_{2r}(\bar{\phi}) - s_{2r}(\phi)](t^r - p^r)} - 1 \right]\,,
\ee
and
\be
{\cal G}(\beta,t){\cal G}(\bar{\beta},p)=e^{\frac{1}{2}\sum_{r=1}^{\infty}
\frac{(-1)^{r+1}}{r}[s_{2r}(\bar{\phi}) + s_{2r}(\phi)](t^r + p^r)}
e^{\frac{1}{2}\sum_{r=1}^{\infty} \frac{(-1)^{r+1}}{r}[s_{2r}(\bar{\phi})
- s_{2r}(\phi)](t^r - p^r)}\,.
\ee
Finally, the second term in (\ref{b}) becomes
\be
2 \frac{t}{p}\frac{t+p}{(1-\frac{t}{p})^2}\left[
{\cal G}(\beta,t){\cal G}(\bar{\beta},p) - {\cal G}(\beta,p)
{\cal G}(\bar{\beta},p)\right]\,.
\ee
We should emphasize that the bosonic part of this expression vanishes.
Combining (C.23) and (C.26) we obtain the final expression for the
generating function $\tilde{\cal A}^{(s)}(t,p)$ in terms of the generating
function ${\cal G}(\beta,p)$:
\be
\tilde{\cal A}^{(s)}(t,p)=
 \frac{4t}{1-\frac{t}{p}}
[t\frac{\pa}{\pa t} -
p\frac{\pa}{\pa p}]{\cal G}(\beta,p){\cal G}(\bar{\beta},t) -
2 \frac{t}{p}\frac{t+p}{(1-\frac{t}{p})^2}\left[
{\cal G}(\beta,t){\cal G}(\bar{\beta},p) - {\cal G}(\beta,p)
{\cal G}(\bar{\beta},p)\right]\,.
\ee
(cf.(C.13)).
Expanding $\tilde{\cal A}^{(s)}(t,p)$ in powers of $t,p$ we arrive at
the explicit expression for the coefficients in front of the second
derivative terms
\bea
\tilde{\cal A}^{(s)}_{\ell,k} &=& 2 \sum_{q\geq 0} \bigg\{ 2(2q + 1 + k -
\ell)\chi_{\ell-q-1}\bar{\chi}_{k+q} - l[\chi_{k+q}\bar{\chi}_{\ell-q-1} -
\bar{\chi}_{k+q}\chi_{\ell-q-1} \cr
\phantom{\Bigg\{}&+& \chi_{\ell+q-1}\bar{\chi}_{k-q} -
\bar{\chi}_{\ell+q-1}\chi_{k-q}]\bigg\}\ .
\eea
(cf.(C.14)) given in (\ref{e5.14}). Making similar calculations for the
coefficients in front of the terms linear in derivatives we come to the 
conclusion
that these coefficients are equal to those given by (5.7) with replacement of
$\si$'s by $\chi$'s (see (5.7) and (5.13)).

\newpage
\def\href#1#2{#2}

\begingroup\raggedright\endgroup

\end{document}